\documentclass[12pt,preprint2]{aastex}
\usepackage{graphics,natbib}

\shorttitle{Galactic Halo velocity-resolved \ion N5 and \ion O6}

\begin{document}
\newcommand{\ts}{\textsuperscript}
\newcommand{\dg}{$^\circ$}
\newcommand{\up}[1]{$\times$10\ts{#1}}
\renewcommand{\O}{\ion{O}{6}}
\newcommand{\N}{\ion{N}{5}}
\newcommand{\C}{\ion{C}{4}}
\newlength{\cw}

\title{\ion O6, \ion N5 and \ion C4 in the Galactic Halo: 
     II. Velocity-Resolved Observations with Hubble and FUSE}
\author{R\'{e}my
Indebetouw\altaffilmark{1} and J. Michael Shull\altaffilmark{2}}

\affil{CASA, Dept. of Astrophysical and Planetary Sciences, University
of Colorado, 389 UCB, Boulder, Colorado 80309-0389 }

\altaffiltext{1}{present address: Astronomy Department, University of
Wisconsin, 475 N. Charter St, Madison, WI 53706 (remy@astro.wisc.edu)} 

\altaffiltext{2}{also at JILA, University of Colorado and National 
Institute of Standards and Technology (mshull@casa.colorado.edu)}

\begin{abstract}

We present a survey of \ion N5 and \ion O6 (and where available \ion
C4) in the Galactic halo, using data from the {\it Far Ultraviolet
Spectroscopic Explorer} (FUSE) and the {\it Hubble Space Telescope}
(HST) along 34 sightlines.  These ions are usually produced in
nonequilibrium processes such as shocks, evaporative interfaces, or
rapidly cooling gas, and thus trace the dynamics of the interstellar
medium.  Searching for global trends in integrated and
velocity-resolved column density ratios, we find large variations in
most measures, with some evidence for a systematic trend of higher
ionization (lower \ion N5/\ion O6 column density ratio) at larger
positive line-of-sight velocities.  The slopes of log[N(\N)/N(\O)] per
unit velocity range from $-0.015$ to $+0.005$, with a mean of $-0.0032
\pm 0.0022$(r)$\pm 0.0014$(sys) dex~(km~s\ts{-1})\ts{-1}.  
We compare
this dataset with models of velocity-resolved high-ion signatures of
several common physical structures.
The dispersion of the ratios, \ion O6/\ion N5/\ion C4, supports the growing
belief that no single model can account for hot halo
gas, and in fact some models predict much stronger trends than are observed.
It is important to understand the signatures of different physical structures 
to interpret specific lines of sight and future global surveys.

\end{abstract}



\newpage

\section{Introduction}

We are interested in understanding how the dynamics of the interstellar 
medium (ISM) of galaxies determines how the energy and matter released 
by stars are redistributed.  In a previous paper \citep{remy04}, hereafter
denoted Paper~I, we investigated theoretical models of the ionization ratios 
of Li-like absorbers, \ion N5, \ion O6, \ion C4, in the Galactic halo.  
These ions are believed to form in nonequilibrium in shocks, evaporative 
interfaces, or rapidly cooling gas, all of which trace the dynamics of the
interstellar medium.  As a useful new diagnostic, in that paper we focused on
{\it velocity-resolved} signatures of several common physical structures:
(1) a cooling Galactic fountain flow that rises, cools, and
    recombines as it returns to the disk;
(2) shocks moving toward the observer;
(3) a conductive interface with the observer located in the hotter gas.
 
In this paper, we present a survey of 34 sightlines observed through
the Galactic halo by the {\it Far Ultraviolet Spectroscopic Explorer}
(FUSE) and the {\it Hubble Space Telescope} (HST) in the Li-like
absorption lines of \ion O6, \ion N5, and sometimes \ion C4.  We
report the integrated column densities of these three ions and their
ratios.  We also focus, as in Paper~I, on the new, and potentially
useful diagnostic provided by the velocity dependence of these ion
ratios.  The observable velocity-ionization trends are weak, because
even strong trends are washed out by the large thermal width of
the gas at different parts of the flow.  Additional confusion results
since the long sightlines almost definitely pass through multiple
structures.  We report an indication of a weak trend of decreasing
N(\N)/N(\O) at more positive velocities.  The distribution of this
slope is broad, with a mean value, $-0.0032\pm0.0022$(r)
$\pm0.0014$(sys)~dex~(km~s\ts{-1})\ts{-1}, displaced somewhat from
zero.

In this paper, \S~\ref{data} describes the data calibration and analysis.  
The basic observational results are described in \S~\ref{results}.  
Section \ref{interpretation} describes models of the dynamical signatures 
of the Li-like ions, and interpretation.

\section{Observations and Data Reduction}
\label{data}

We observe the relative distribution of Li-like ions in the
interstellar medium using the FUSE satellite and the Space Telescope
Imaging Spectrograph (STIS) and Goddard High Resolution Spectrograph
(GHRS) on HST.  The FUSE satellite consists of four coaligned Rowland
circle spectrographs which redundantly cover the spectral range
910-1189~\AA\ at $R\sim$20,000 resolution
\citep{fuse1,sahnow00}.  The STIS instrument 
offers far-ultraviolet spectroscopy in first-order and echelle modes. 
Spectroscopic modes used in this study are the G140M first-order
grating (55~\AA\ in the range 1140-1740~\AA, $R\sim$15,000) and the
E140M echelle (1150-1710~\AA, $R\sim$50,000) \citep{stis,stisweb}.
The GHRS was the precursor UV spectrograph to STIS on the HST, and
offered similar capabilities without the imaging capability of STIS.
Data used here were obtained with the G160M first-order grating
\citep{ghrs,ghrsweb}.

The data analysis consists of instrument-dependent calibration,
followed by a common continuum and wavelength verification process.
The corrections which are unique to a given bandpass are made (e.g.,
assessment of H$_{\rm 2}$ contamination which is only relevant to the
\O\ band), and finally the two or three bandpasses are combined to
study the velocity-resolved \C/\N/\O\ column density ratios. Table
\ref{datasets} shows the sightlines, with the instrument and dataset
name for each (multi-exposure) observation, the total (nighttime for
FUSE) exposure time, observation date, the principal investigator, and
for HST instruments, the proposal ID number.  All of the FUSE data are
guaranteed time observations (GTO), and the GTO project can be
identified by the first 3 numbers of the dataset name.  STIS and GHRS
first-order spectra only cover the \C\ or the \N\ line; for those
datasets the relevant ion is identified in the instrument column of
the table.  As described below, the GHRS wavelength scale can be
improved using auxiliary datasets, and the ones used in this project
are also listed.

\subsection{Basic Calibration and Instrument-Specific Issues}

\subsubsection{Far Ultraviolet Spectroscopic Explorer}

Most FUSE data issues are adequately resolved by the FUSE data
pipeline \citep[version 1.1 is described in][all data used here were
re-reduced using version 1.8.7 and spot checked against
2.0.5]{fusedata}.  The pipeline screens for basic satellite location
and orientation problems -- South Atlantic Anomaly avoidance, earth
avoidance, etc. \citep[see][]{3c273}.  Particular to FUSE are the
thermal motions of the grating and telescope with orbital phase and
spacecraft orientation with respect to the satellite orbit plane and the sun.
Thermally induced grating motion is well understood and removed by the
pipeline.  Telescope motion is removed in the LiF1 channel because the
spacecraft points using that optical path.  A point source can move in
the large apertures of the second lithium fluoride channel LiF2 and
both silicon carbide channels SiC1 and SiC2, and the spectral
resolution and instrument throughput can be degraded.  The detector
characteristics are sufficiently different that it is inadvisable to
directly combine data from different segments. Therefore, for \O, the
LiF1A segment are used for quantitative measurements, and the LiF2B
data used for feature confirmation in cases of low signal-to-noise.

The FUSE data for each target were obtained from the archive in raw
form, re-reduced and combined into single spectrum for each detector
segment, with the following steps. A target may be observed several
times, resulting in several {\it datasets}, each consisting of
multiple {\it exposures}.  All exposures were screened for bursts, and
the remaining data were recalibrated using the FUSE pipeline
(calibrated data was available from the archive but from earlier
versions of the pipeline).  The pipeline removes the time-dependent
spectral motion from the grating motion and the orbital Doppler shift,
removes detector background, extracts a 1-dimensional spectrum, and
provides a flux calibration and wavelength solution.  This
recalibration also allows us to select only photons received during
spacecraft night.  Although not absolutely necessary for analysis of
the \O\ line in isolation, there are known artifacts in daytime data.
For simplicity in this study, only nighttime data were used.

The FUSE pipeline wavelength solution at the time of this analysis was
questionable, especially in its total offset (the relative scale is
somewhat more trustworthy).  For example, the heliocentric-to-LSR
conversion was applied with incorrect sign in v1.8.7.  Multiple
datasets were cross-correlated to find their relative wavelength
offsets before combining, and the absolute wavelength scale was
corrected using low-ionization interstellar lines as described below.
Several datasets were recalibrated with v2.0.5, and no significant
differences were found in the spectral region considered here, aside
from the overall velocity offset, which had been corrected anyway.

\subsubsection{Space Telescope Imaging Spectrograph}

STIS echelle data suffer from cross-order scatter, which degrades the
spectral purity and increases the background in the extracted spectra.
In fact, the standard STIS pipeline uses a linear interpolation of the
count rate between orders to subtract the background when extracting a
one-dimensional spectrum from the two-dimensional image.  A better
background algorithm, which models the two-dimensional scatter
iteratively with the science spectrum, can be included in the STIS
pipeline.  All STIS echelle datasets were recalibrated through the
pipeline with this additional algorithm.  The pipeline performs
Doppler correction, flat and dark fielding, flux calibration, and
wavelength solution.  First-order STIS data were deemed acceptable in
post-pipeline calibrated form from the archive.

\subsubsection{Goddard High Resolution Spectrograph}

All data were recalibrated using the most recent version of the GHRS
pipeline and calibration files.  Changes especially in the latter make
this recalibration preferable to using archived calibrated data.
During the recalibration process, the GHRS wavelength scale was
improved using calibration lamp and SPYBAL exposures taken immediately
before or after the science exposure.  The pipeline combines
substepped exposures, performs flat and dark fielding, and provides
flux calibration and a wavelength solution.  Much of the data used
here were taken as part of surveys for low-redshift Ly$\alpha$
absorption \citep{penton2,penton1,penton3}.  The basic data reduction steps
were repeated here, but following the same procedure described in
these papers, to which the reader is referred 
for further discussion of the data characteristics.

\begin{deluxetable}{lcccccllllll}
\renewcommand{\N}{NV}
\renewcommand{\O}{OVI}
\renewcommand{\C}{CIV}
\tabletypesize{\scriptsize}
\setlength{\tabcolsep}{1mm}
\tablecaption{\label{datasets}Datasets}
\tablehead{
\colhead{Target} &
\colhead{Instrument\tablenotemark{a}} &
\colhead{\parbox[b]{5em}{Diagnostic\\line}} &
\colhead{dataset(s)} &
\colhead{aux dataset(s)} &
\colhead{\parbox[b]{3em}{$t_{\rm{exp}}$\\(ksec)}} &
\colhead{\parbox[b]{3em}{obsv\\date}} &
\colhead{PI/PEP ID}
}
\startdata
ESO 141-G55
& FUSE & \O & i9040104          &               & 11.1  & 9/99  & Savage\\* 
& GHRS & \C & z3i7010dt         & z3i7010ct     & 4.6   & 10/96 & Savage/6451\\*
& GHRS & \C & z3i7010bt         & z3i7010ct     & 6.5   & 10/96 & Savage/6451\\*
& GHRS & \N & z3i70105t         & z3i7010ct     & 9.8   & 10/96 & Savage/6451\\*
& GHRS & \N & z3e70204t         & z3e70203t     & 7.8   & 8/96  & Stocke/6593\\*
& GHRS & \N & z3e70206t         & z3e70205t     & 7.8   & 8/96  & Stocke/6593\\
\tableline
Fairall 9
& FUSE & \O & p1010601          &               & 4.0   & 7/00  & Savage\\* 
& GHRS & \N & z3e70406t         & z3e70405t     & 6.9   & 8/96  & Stocke/6593\\*
& GHRS & \N & z3e70404m         & z3e70403t     & 14.4  & 8/96  & Stocke/6593\\*
& GHRS & \C & z26o020bt         & z26o020at     & 8.1   & 4/94  & Savage/5300\\*
& GHRS & \N & z26o0208n         & z26o0207t,at  & 8.1   & 4/94  & Savage/5300\\
\tableline
H 1821+643
& FUSE & \O & p1016402          &               & 47.3  & 7/00  & Savage\\* 
& E140M&\C,\N& o5e704010-e0      &               & 24.5  & 3/00  & Jenkins/8165\\*
& E140M&\C,\N& o5e703010-e0      &               & 24.5  & 7/99  & Jenkins/8165\\*
& G140M& \C & o5fe01010         &               & 3.2   & 6/99  & Stocke/8182\\*
& GHRS & \C & z27n010bn         & z27n010at,ct  & 16.1  & 4/94  & Lu/5299\\*
& GHRS & \N & z27n0108m         & z27n0107t,9t  & 10.4  & 4/94  & Lu/5299\\*
& GHRS & \N & z15f0208m         & z15f0207t     & 7.3   & 4/93  & Savage/4094\\
\tableline
Mrk 110
& FUSE & \O & p1071302  &               & 7.7  & 2/01  & Moos\\* 
& G140M& \N & o4n302010 &               & 1.1   & 5/99  & C\^ot\'e/7295\\*
& G140M& \N & o4n352010 &               & 1.1   & 10/99 & C\^ot\'e/7295\\
\tableline
Mrk 116 
& FUSE & \O & p1080901          &               & 38.2  & 2/01  & Moos\\* 
& GHRS & \N & z3ig020bt         & z3ig020at     & 8.3   & 12/96 & Lequeux/6710\\*
& GHRS & \N & z3ig0208m         & z3ig0207t     & 6.5   & 12/96 & Lequeux/6710\\*
& GHRS & \N & z3ig0205t         & z3ig0204t     & 6.5   & 12/96 & Lequeux/6710\\*
& GHRS & \N & z3ig010bt         & z3ig010at     & 8.3   & 11/96 & Lequeux/6710\\*
& GHRS & \N & z3ig0108m         & z3ig0107t     & 6.5   & 11/96 & Lequeux/6710\\*
& GHRS & \N & z3ig0105t         & z3ig0104t     & 6.5   & 11/96 & Lequeux/6710\\*
& GHRS & \N & z3ig0305t         & z3ig0304t     & 3.5   & 11/96 & Lequeux/6710\\
\tableline
Mrk 279
& FUSE & \O & p1080303          &               & 17.8  & 12/99 & Moos\\* 
& FUSE & \O & p1080304          &               & 3.2  & 1/00  & Moos\\* 
& GHRS & \N & z3e70304t         & z3e70303t,5t  & 6.1   & 1/97  & Stocke/6593\\*
& GHRS & \N & z3e70306t         & z3e70305t,7t  & 6.1   & 1/97  & Stocke/6593\\*
& GHRS & \N & z3e70308t         & z3e70307t,9t  & 6.1   & 1/97  & Stocke/6593\\*
& GHRS & \N & z3e7030at         & z3e70309t     & 1.5   & 1/97  & Stocke/6593\\
\tableline
Mrk 290
& FUSE & \O & p1072901          &               & 6.2  & 3/00  & Moos\\* 
& GHRS & \N & z3kh0104t         & z3kh0103t     & 1.8   & 1/97  & Wakker/6590\\*
& GHRS & \N & z3kh0105t         & z3kh0103t,6t  & 2.8   & 1/97  & Wakker/6590\\*
& GHRS & \N & z3kh0107t         & z3kh0106t     & 2.5   & 1/97  & Wakker/6590\\
\tableline
Mrk 335
& FUSE & \O & p1010203          &               & 28.9 & 12/99 & Savage\\* 
& FUSE & \O & p1010204          &               & 46.2  & 11/00 & Savage\\* 
& GHRS & \N & z1a60304n         & z1a60303t     & 14.8  & 9/93  & Stocke/3584\\
\tableline
Mrk 421
& FUSE & \O & p1012901          &               & 8.2  & 12/00 & Savage\\* 
& GHRS & \N & z2ia0104t         & z2ia0103t     & 15.7  & 2/95  & Stocke/5715\\
\tableline
Mrk 478
& FUSE & \O & p1110909          &               & 9.5  & 1/29  & Kriss\\* 
& G140M& \N & o4ec14040,50,60   &               & 7.6   & 3/99  & Stocke/7345\\
\tableline
Mrk 509
& FUSE & \O & x0170101,2        &               & 42.2  & 11/99 & Sembach\\* 
& GHRS & \N & z3e70704t         & z3e70703t     & 4.7   & 10/96 & Stocke/6593\\*
& GHRS & \C & z26o0108t         & z26o0107t,9t  & 8.1   & 8/94  & Savage/5300\\*
& GHRS & \N & z1790208m         & z1790207t,9m  & 6.9   & 4/93  & Savage/3463\\
\tableline
Mrk 771 
& FUSE & \O & p1072301          &               & 6.1   & 1/01  & Moos\\* 
& G140M& \N & o4ec07030,40      &               & 5.8   & 7/99  & Stocke/7345\\*
& G140M& \N & o4n305010         &               & 2.0   & 5/99  & C\^ot\'e/7295\\
\tableline
Mrk 817
& FUSE & \O & p1080401,2        &               & 11.5  & 2/00  & Moos\\* 
& FUSE & \O & p1080404          &               & 16.4  & 2/01  & Moos\\* 
& GHRS & \N & z3e70104t         & z3e70103t     & 9.5   & 1/97  & Stocke/6593\\*
& GHRS & \N & z3e70106t         & z3e70105t     & 8.8   & 1/97  & Stocke/6593\\*
& GHRS & \N & z3e70108t         & z3e70107t     & 8.5   & 1/97  & Stocke/6593\\
\tableline
Mrk 876
& FUSE & \O & p1073101          &               & 20.9  & 10/99 & Moos\\* 
& G140M& \N & o4n308010         &               & 2.3   & 9/98  & Cote/7295\\
\tableline
Mrk 926
& FUSE & \O & p1074001          &               & 4.0  & 6/00  & Moos\\* 
& G140M& \N & o4ec12020,10      &               & 3.9   & 7/99  & Stocke/7345\\
\tableline
Mrk 1095 
& FUSE & \O & p1011201,2        &               & 13.9  & 11/00 & Savage\\* 
& FUSE & \O & p1011203          &               & 16.9  & 12/00 & Savage\\* 
& GHRS & \N & z3e7060at         & z3e70609t     & 2.6   & 12/96 & Stocke/6593\\*
& GHRS & \N & z3e70608m         & z3e70607t     & 6.1   & 12/96 & Stocke/6593\\*
& GHRS & \N & z3e70606t         & z3e70605t     & 6.1   & 12/96 & Stocke/6593\\*
& GHRS & \N & z3e70604t         & z3e70603t     & 6.1   & 12/96 & Stocke/6593\\
\tableline
NGC 3783
& FUSE & \O & p1013301          &               & 23.4  & 2/00  & Savage\\* 
& E140M&\C\N& o57b01010,20,30   &               & 7.6   & 2/00  & Kraemer/8029\\*
& GHRS & \C & z2nf0107t         & z2nf0106t,9t  & 2.2   & 4/95  & Maran/5724\\*
& GHRS & \C & z2nf0108t         & z2nf0106t,9t  & 2.2   & 4/95  & Maran/5724\\*
& GHRS & \C & z1av0107t         & z1av0106t,9t  & 2.4   & 1/94  & Maran/3936\\*
& GHRS & \C & z1av0108t         & z1av0106t,9t  & 2.4   & 1/94  & Maran/3936\\*
& GHRS & \C & z11q0107m         & z11q0106m,8m  & 2.2   & 2/93  & Maran/1160\\*
& GHRS & \C & z11q0109m         & z11q0108m,am  & 2.0   & 2/93  & Maran/1160\\*
& GHRS & \N & z1790308m         & z1790307t,9m  & 8.1   & 2/93  & Savage/3463\\
\tableline
NGC 4151
& FUSE & \O & p1110505  &                       & 9.1  & 3/00  & Kriss\\* 
& E140M&\C,\N& o61l01010,20      &               & 9.4   & 3/00  & Kriss/8608\\*
& E140M&\C,\N& o57801030,40      &               & 5.5   & 7/99  & Hutchings/8019\\*
& GHRS & \C & z2bs0405p-ap      & z2bs0404p     & 9.2   & 3/96  & Weymann/5169\\*
& GHRS & \C & z2bs0205t-8t,at,bt & z2bs0204t,9t & 9.2 & 10/94 & Weymann/5169\\*
& GHRS & \C & z1li0205t-8t,at,bt & z1li0204t,9t & 9.2 & 1/94 & Weymann/4880\\*
& GHRS & \C & z0yd0209t         & z0yd0207t,8t,at & 1.4 & 7/92  & Weymann/1141\\*
& GHRS & \C & z0yd020bm         & z0yd020at     & 1.4   & 7/92  & Weymann/1141\\*
& GHRS & \C & z0yd020cm         & z0yd020dm     & 1.4   & 7/92  & Weymann/1141\\*
& GHRS & \C & z0yd020em         & z0yd020dm     & 1.4   & 7/92  & Weymann/1141\\*
& GHRS & \C & z0yd020fm         & z0yd020gm     & 1.4   & 7/92  & Weymann/1141\\*
& GHRS & \C & z0yd020hm         & z0yd020gm     & 1.4   & 7/92  & Weymann/1141\\
\tableline
NGC 5548
& FUSE & \O & p1014601          &               & 20.9  & 6/00  & Savage\\* 
& E140M&\C,\N& o4ll010d0,c0      &               & 4.7   & 3/98  & Kraemer/7572\\*
& GHRS & \C & z2ws020at,9t      & z2ws0208t     & 9.9   & 2/96  & Savage/5889\\*
& GHRS & \N & z2ws0207t         & z2ws0205t,6t  & 4.6   & 2/96  & Savage/5889\\
\tableline
PG 0804+761
& FUSE & \O & p1011901          &               & 5.9  & 10/99 & Savage\\* 
& FUSE & \O & p1011903          &               & 1.5  & 1/00  & Savage\\* 
& G140M& \N & o4ec06010         &               & 4.9   & 2/98  & Stocke/7345\\*
& G140M& \N & o4n301010         &               & 2.4   & 2/98  & C\^ot\'e/7295\\
\tableline
PG 0953+414
& FUSE & \O & p1012201          &               & 26.7  & 5/00  & Savage\\* 
& FUSE & \O & p1012202  &                       & 15.9  & 12/99 & Savage\\* 
& E140M&\C\N& o4x001010,20,2010 &               & 244.8 & 12/98 & Savage/7747\\
\tableline
PG 1116+215
& FUSE & \O & p1013101          &               & 11.0  & 4/00  & Savage\\* 
& FUSE & \O & p1013103,4,5      &               & 43.6  & 4/01  & Savage\\* 
& E140M&\C,\N& o5a301010,20,2010,20 &            & 199.2 & 5/00  & Sembach/8097\\*
& G140M& \N & o4ec04010,20      &               & 2.6   & 5/00  & Stocke/7345\\
\tableline
PG 1211+143
& FUSE & \O & p1072001          &               & 52.2  & 4/00  & Moos\\* 
& G140M& \N & o4ec08040,30      &               & 5.8   & 7/99  & Stocke/7345\\
\tableline
PG 1259+593
& FUSE & \O & p1080101          &               & 14.0  & 2/00  & Moos\\* 
& FUSE & \O & p1080102          &               & 0.5  & 12/00 & Moos\\* 
& FUSE & \O & p1080103          &               & 35.4  & 1/01  & Moos\\* 
& FUSE & \O & p1080104--9       &               & 303.5 & 3/01  & Moos\\* 
& G140M& \N & o4n307020         &               & 1.8   & 3/99  & C\^ot\'e/7295\\
\tableline
PG 1351+640
& FUSE & \O & p1072501          &               & 25.1  & 1/00  & Moos\\* 
& G140M& \N & o4ec54010         &               & 8.5   & 8/00  & Stocke/7345\\
\tableline
PKS 2005-489
& FUSE & \O & p1073801          &               & 7.7  & 8/00  & Moos\\* 
& G140M& \N & o4ec09040,30      &               & 6.1   &       & Stocke/7345\\
\tableline
PKS 2155-304
& FUSE & \O & p1080701          &               & 17.7  & 10/99 & Moos\\* 
& FUSE & \O & p1080705          &               & 34.8  & 10/99 & Moos\\* 
& E140M&\C,\N& o5by02010,20      &               & 14.3  & 9/00  & Shull/8125\\*
& E140M&\C,\N& o5by01010,20      &               & 14.3  & 11/99 & Shull/8125\\*
& GHRS & \C & z2ws0107p,8p      & z2ws0106p     & 6.7   & 10/95 & Savage/5889\\*
& GHRS & \N & z1aw0106t,7t,8m   & z1aw0105t     & 5.3   & 5/93  & Bogess/3965\\
\tableline
Q1230.8+0115
& FUSE & \O & p1019001          &               & 4.0   & 6/00  & Savage\\*
& E140M&\C,\N& o56a01010,20,2010,20 &            & 27.2  & 1/99  & Rauch/7737\\*
& GHRS & \N & z3cj0105t         & z3cj0104t     & 5.9   & 7/96  & Rauch/6410\\*
& GHRS & \N & z3cj0108t         & z3cj0107t     & 5.9   & 7/96  & Rauch/6410\\
\tableline
3C273
& FUSE & \O & p1013501          &               & 29.8  & 4/00  & Savage\\* 
& GHRS & \C & z1d00109t         & z1d00108t,at  & 1.1   & 12/93 & Weymann/4883 \\*
& GHRS & \C & z1d0010bt-dt      & z1d0010at,et  & 3.3   & 12/93 & Weymann/4883 \\*
& GHRS & \N & z1760105t         & z1760103t,4t,6t & 1.2 & 11/93 & Weymann/3951 \\*
& GHRS & \N & z1760107t,8t      & z1760106t     & 2.4   & 11/93 & Weymann/3951 \\*
& GHRS & \N & z0gu010om         & z0gu010nm,pm  & 2.9   & 2/91  & Weymann/1140 \\*
& GHRS & \C & z0gu010cm         & z0gu010am,bm,dm & 1.4 & 2/91  & Weymann/1140 \\
\tableline
3C351
& FUSE & \O & q1060101          &               & 17.0  & 10/99 & le Brun \\* 
& FUSE & \O & q1060102          &               &  3.1  & 2/00  & le Brun \\* 
& FUSE & \O & p1080801          &               & 53.5  & 5/01  & Moos \\* 
& E140M&\C,\N& o57902010-80 &                & 18.6  & 2/00  & Jenkins/8015\\*
& E140M&\C,\N& o57901010-80,3020-60 &       & 34.2  & 6/99  & Jenkins/8015\\
\tableline
Ton S180
& FUSE & \O & p1010502          &               & 12.3  & 12/99 & Savage\\* 
& G140M& \N & o4ec02010,20      &               &  4.1  & 7/99  & Stocke/7345\\
\tableline
VII Zw 118
& FUSE & \O & p1011604          &               & 15.1  & 10/99 & Savage\\* 
& FUSE & \O & p1011605          &               & 9.8  & 11/99 & Savage\\* 
& FUSE & \O & p1011606          &               & 7.2   & 1/00  & Savage\\* 
& G140M& \N & o4ec13010         &               & 9.5   & 11/97 & Stocke/7345
\enddata
\renewcommand{\O}{\ion{O}{6}}
\renewcommand{\N}{\ion{N}{5}}
\renewcommand{\C}{\ion{C}{4}}
\tablenotetext{a}{``E140M'' and ``G140M'' refer to echelle and first-order modes
of STIS, whereas ``GHRS'' refers in all cases to the G160M first-order mode of that
instrument.}
\end{deluxetable}

\clearpage

\section{Observational Results}
\label{results}

\subsection{Summary of Observations}

The data are of widely varying quality, but show good agreement with
previous measurements.  The dataset as a whole is well-suited to the
study of global trends.  Unfortunately, the values of N(\N)/N(\O) do
not seem to favor any of the production mechanisms modeled in the
literature.  We find that there may be a weak trend for lower
N(\N)/N(\O) at more positive velocities, which can be thought of as a
higher ionization state at more positive velocity (redshifted) gas.
Previous authors have discussed such a signature as an offset in mean
velocity between the Li-like ion absorption.  Although this is an
acceptable way to think of the situation, it implies that the
absorption profiles are the same, which is not always true.

Table \ref{targets} summarizes the data for the 34 lines of sight.
The total (integrated) column densities for the \O, \N, and, where
available, \C\ absorption lines are listed along with the velocity
range [$v_-$,$v_+$] over which the line was integrated.  
For \N\ and \C, the column densities calculated from the two lines of
the doublet agree to within errors, and there is no evidence that a
saturation correction for instrumental smearing is required. Thus, the
weighted mean column density is used in subsequent analysis.
Also listed are the 3$\sigma$ upper limit equivalent widths for
\ion{Cl}{1} and H$_{\rm 2}$ lines, or in cases where H$_{\rm 2}$ 
absorption was removed from the \O\ line, the equivalent width is
listed.  \N\ is typically weak, but many sightlines have reliable
detections.  
Table \ref{targets} also lists the slope of a line fitted to the log
of the column density ratio N(\N)/N(\O), a procedure discussed in the
next section.
Selected characteristics of some sightlines are noted in
Appendix~\ref{appendix}.

\newcommand{\tn}[1]{{\tablenotemark{#1}}}
\begin{deluxetable}{lrrccccccccccccccccc}
\rotate
\renewcommand{\N}{NV}
\renewcommand{\O}{OVI}
\renewcommand{\C}{CIV}
\tabletypesize{\tiny}
\tablewidth{0cm}
\setlength{\tabcolsep}{1.0mm}
\tablecaption{\label{targets} Sightlines}
\tablehead{
\colhead{Target} & \colhead{{l}} & \colhead{b} &
\colhead{\parbox{2.em}{\begin{center}EW\\ClI\\(m\AA)\end{center}}} &
\colhead{\parbox{3.em}{\begin{center}EW\\H$_2\lambda$1031\\(m\AA)\end{center}}} &
\colhead{\parbox{3.em}{\begin{center}EW\\H$_2\lambda$1032\\(m\AA)\end{center}}}
 &
\multicolumn{2}{c}{\parbox{4. em}{\begin{center}~\\v$_\pm$\\(km~s\ts{-1})\end{center}}} &
\colhead{\parbox{2. em}{\begin{center}log~N\\\O\\$\lambda1032$\end{center}}} &
\colhead{\parbox{2. em}{\begin{center}\ \\$\delta$\\log~N~\end{center}}} &
\colhead{\parbox{2. em}{\begin{center}log~N\\\N\\$\lambda1239$\end{center}}} &
\colhead{\parbox{2. em}{\begin{center}\ \\$\delta$\\log~N~\end{center}}} &
\colhead{\parbox{2. em}{\begin{center}log~N\\\N\\$\lambda1242$\end{center}}} &
\colhead{\parbox{2. em}{\begin{center}\ \\$\delta$\\log~N~\end{center}}} &
\colhead{\parbox{2. em}{\begin{center}log~N\\\C\\$\lambda1548$\end{center}}} &
\colhead{\parbox{2. em}{\begin{center}\ \\$\delta$\\log~N~\end{center}}} &
\colhead{\parbox{2. em}{\begin{center}log~N\\\C\\$\lambda1551$\end{center}}} &
\colhead{\parbox{2. em}{\begin{center}\ \\$\delta$\\log~N~\end{center}}} &
\colhead{\parbox{2. em}{\begin{center}slope\\\underline{\N}\\\O\end{center}}} &
\colhead{\parbox{2. em}{\begin{center}$\delta$slope\\\underline{\N}\\\O\end{center}}} 
}
\newcommand{\pv}{~~~~}
\newcommand{\pw}{~~~}
\startdata
Parkes 2155-304 &  17.7 & -52.3 & $<$  5.6 &   \pv$<$~~9      &  \pw$<$~~6 & -300 & 150      & 14.52 & .07 &  13.53 & .35 &  13.25      &$>$1 & 14.21 & .06 & 14.16 & .13 & -.0005 & .0003 \\
NGC 5548        &  32.0 &  70.5 & $<$ 36   & ~39$\pm$~14      &  \pw$<$~27 & -150 & 100      & 14.52 & .11 &  13.73 & .17 &  13.50      & .65 & 14.43 & .05 & 14.45 & .08 & -.0034 & .0003 \\
Markarian 509   &  36.0 & -29.9 & $<$ 11   & ~81$\pm$~12\tn{b}& 21$\pm$~32 & -100 & 115      & 14.68 & .05 &  13.86 & .14 &  13.73      & .33 & 14.39 & .05 & 13.46 & .07 & +.0005 & .0001 \\
Markarian 478   &  59.2 &  65.0 & $<$ 32   & ~25$\pm$~22      &  \pw$<$~26 & -150 & 100      & 14.50 & .12 &  13.47 & .34 &  13.53      & .60 &  -    &     &  -    &     & -.0039 & .0006 \\
Markarian 1513  &  63.7 & -29.1 & $<$ 43   & ~73$\pm$~20      &  \pw$<$~78 & -100 & 100      & 14.33 & .17 &  13.19 & .66 &  13.69      & .42 &  -    &     &  -    &     & -.0035 & .0020 \\
Markarian 926   &  64.1 & -58.8 & $<$225   &   \pv$<$257      &  \pw$<$140 & ~-65 & 100      & 14.26 & .65 &  13.00 & .97 &$<$14.3      &     &  -    &     &  -    &     &    -   &   -   \\
Markarian 290   &  91.5 &  48.0 & $<$ 77   &   \pv$<$122      &  \pw$<$~42 & -200 & 200      & 14.48 & .22 &  13.56 & .42 &  13.28      &1.   &  -    &     &  -    &     & +.0016 & .0016 \\
H 1821+643      &  94.0 &  27.4 & $<$ 25   & ~35$\pm$~10      &  \pw$<$~13 & -150 & 100\tn{c}& 14.54 & .08 &  14.09 & .10 &  14.20      & .15 & 14.39 & .04 & 14.40 & .07 & +.0002 & .0003 \\
Markarian 876   &  98.3 &  40.4 & $<$ 12   & ~85$\pm$~10      & 24$\pm$~22 & -200 & 150\tn{c}& 14.60 & .08 &  13.58 & .35 &  13.75      & .46 &  -    &     &  -    &     & -.0024 & .0002 \\
Markarian 817   & 100.3 &  53.5 & $<$  9.5 &   \pv$<$~11      &  \pw$<$~10 & -200 & 150\tn{c}& 14.56 & .04 &  13.87 & .10 &  12.93      &$>$1 &  -    &     &  -    &     & -.0007 & .0001 \\
Markarian 335   & 108.8 & -41.4 & $<$  7.0 & ~73$\pm$~~6      & 22$\pm$~14 & -100 & 100      & 14.06 & .16 &$<$13.5 & ~~~ &  13.10      & .50 &  -    &     &  -    &     &    -   &   -   \\
PG 1351+640     & 111.9 &  52.0 & $<$ 27   & ~62$\pm$~20      &  \pw$<$~29 & -175 & 175      & 14.57 & .12 &$<$14.2 & ~~~ &  12.80      &$>$1 &  -    &     &  -    &     &    -   &   -   \\
Markarian 279   & 115.0 &  46.9 & $<$ 11   &   \pv$<$~13      &  \pw$<$~18 & -200 & 100      & 14.52 & .08 &  13.74 & .15 &  13.56\tn{d}& .40 &  -    &     &  -    &     & -.0025 & .0001 \\
PG 1259+593     & 120.6 &  58.1 & $<$ 14   & ~10$\pm$~16      &  \pw$<$~~5 & -175 & 200\tn{c}& 14.35 & .06 &  13.25 &1.~~ &  14.22\tn{d}& .30 &  -    &     &  -    &     & -.0038 & .0051 \\
Markarian 1502  & 123.8 & -50.2 & $<$ 70   & ~60$\pm$~60      & 51$\pm$~40 & ~-50 & ~85      & 14.00 & .27 &  13.25 & .60 &$<$13.8      &     &  -    &     &  -    &     & -.0021 & .0075 \\
PG 0804+761     & 138.3 &  31.0 & $<$ 16   & 104$\pm$~14      & 22$\pm$~26 & -125 & 100      & 14.48 & .09 &  13.34 & .27 &$<$13.9      &     &  -    &     &  -    &     & -.0109 & .0004 \\
Ton S180        & 139.0 & -85.1 & $<$ 25   &   \pv$<$~23      &  \pw$<$~16 & ~-95 & 100      & 14.39 & .10 &  13.10 & .61 &  12.90      &$>$1 &  -    &     &  -    &     & -.0053 & .0008 \\
Zwicky VII 118  & 151.4 &  26.0 & $<$ 19   & ~73$\pm$~~8      & 32$\pm$~20 & -100 & ~75      & 14.21 & .13 &  12.30 &$>$1 &  13.40      & .86 &  -    &     &  -    &     &    -   &   -   \\
NGC 4151        & 155.1 &  75.1 & $<$ 29   & 108$\pm$~20\tn{b}&  \pw$<$~15 & -100 & 100\tn{c}& 14.12 & .09 &  12.96 &1.~~ &  -\tn{d}    &     & 14.01 & .06 & 13.75 & .34 & -.0021 & .0027 \\
Markarian 116   & 160.5 &  44.8 & $<$ 21   & ~55$\pm$~14      &  \pw$<$~78 & -150 & 100      & 14.40 & .22 &  13.00 &$>$1 &  13.60      & .58 &  -    &     &  -    &     &    -   &   -   \\
Markarian 110   & 165.0 &  44.4 & $<$288   & ~50$\pm$100      &  \pw$<$140 & -100 & 100      & 14.30 & .32 &  13.28 & .83 &$<$14.4      &     &  -    &     &  -    &     & +.0092 & .0086 \\
PG 0953+414     & 179.8 &  51.7 & $<$ 13   &   \pv$<$~~9      &  \pw$<$~~9 & -125 & 250\tn{c}& 14.50 & .08 &  13.26 & .87 &  13.77      & .53 & 13.97 & .13 & 13.64 & .50 & +.0068 & .0036 \\
Markarian 421   & 179.8 &  65.0 & $<$ 18   & ~31$\pm$~46      &  \pw$<$~16 & -150 & 200\tn{c}& 14.44 & .12 &  13.30 & .48 &  12.30\tn{d}&1.   &  -    &     &  -    &     & -.0067 & .0036 \\
NGC 985         & 180.8 & -59.5 & $<$ 36   & ~34$\pm$~10      &  \pw$<$~15 & -100 & 100      & 14.32 & .12 &  13.41 & .32 &  13.79      & .25 &  -    &     &  -    &     & -.0022 & .0002 \\
Markarian 1095  & 201.7 & -21.1 & $<$ 12   & 115$\pm$~10      &  \pw$<$~20 & ~-50 & ~50      & 13.81 & .25 &  13.32 & .28 &  13.68      & .23 &  -    &     &  -    &     & -.0156 & .0007 \\
PG 1116+215     & 223.4 &  68.2 & $<$  8   & ~15$\pm$~20      &  \pw$<$~~9 & ~-50 & 350      & 14.59 & .10 &  13.89 & .27 &  13.73      & .74 & 14.31 & .10 & 14.35 & .14 & -.0000 & .0003 \\
PG 1211+143     & 267.6 &  74.3 & $<$ 11   & ~49$\pm$~~6      &  \pw$<$~11 & -100 & 100      & 14.08 & .17 &  13.64 & .18 &  14.01\tn{d}& .15 &  -    &     &  -    &     & -.0035 & .0004 \\
PG 1229+204     & 269.4 &  81.7 & $<$137   &   \pv$<$~58      &  \pw$<$~54 & ~-50 & 150      & 14.57 & .13 &  13.44 & .31 &  13.72      & .30 &  -    &     &  -    &     & -.0020 & .0014 \\
NGC 3783        & 287.5 &  23.0 & $<$ 17   & 149$\pm$~19\tn{b}& 95$\pm$165 & -100 & 100      & 14.52 & .09 &  13.23 & .72 &  12.70      &$>$1 & 14.30 & .06 & 14.26 & .10 & -.0041 & .0006 \\
3C273           & 290.0 &  64.4 & $<$  4.5 & ~27$\pm$~~6      &  \pw$<$~~8 & -110 & 275      & 14.78 & .04 &  13.94 & .11 &  13.88      & .25 & 14.47 & .03 & 14.50 & .05 & -.0004 & .0001 \\
Fairall 9       & 295.1 & -57.8 & $<$230   &   \pv$<$~90      &  \pw$<$~83 & -100 & 250      & 14.51 & .20 &  13.40 & .36 &  13.78\tn{d}& .27 & 14.19 & .10 & 14.43 & .10 & -.0011 & .0047 \\
ESO 141-G55     & 338.2 & -26.7 & $<$ 11   & ~43$\pm$~12      & 27$\pm$~16 & -100 & 100      & 14.49 & .08 &  13.68 & .15 &  13.77      & .23 & 14.41 & .04 & 14.55 & .05 & -.0020 & .0001 \\
Markarian 1383  & 349.2 &  55.1 & $<$ 19   & ~15$\pm$~30      &  \pw$<$~16 & -100 & 160      & 14.48 & .08 &  13.85 & .12 &  13.80      & .25 &  -    &     &  -    &     & -.0009 & .0001 \\
Parkes 2005-489 & 350.4 & -32.6 & $<$ 18   &   \pv$<$~25      &  \pw$<$~23 & -100 & 250      & 14.88 & .05 &  14.05 & .07 &  13.94      & .17 &  -    &     &  -    &     & +.0031 & .0002 \\
\enddata
\tablenotetext{a}{Upper limit equivalent widths are 3$\sigma$, and
        errors are 2$\sigma$.}
\tablenotetext{b}{Complex structure in H$_2$: multiple components removed.}
\tablenotetext{c}{In a few sightlines, the range over which the line
ratio $\log$N(\N)/$\log$N(\O) had to be restricted because of low
signal-to-noise in the \N line. The particular sightlines and their
restricted ranges are: H~1821+643, [-120,0]; Markarian~421, [-60,50];
Markarian~817, [-200,50]; Markarian~876, [-40,100]; NGC~4151,
[-25,60]; PG~0953+414, [-50,10]; PG~1259+593, [-30,50].}
\tablenotetext{d}{Line contaminated.}

\end{deluxetable}

\begin{figure}
\plotone{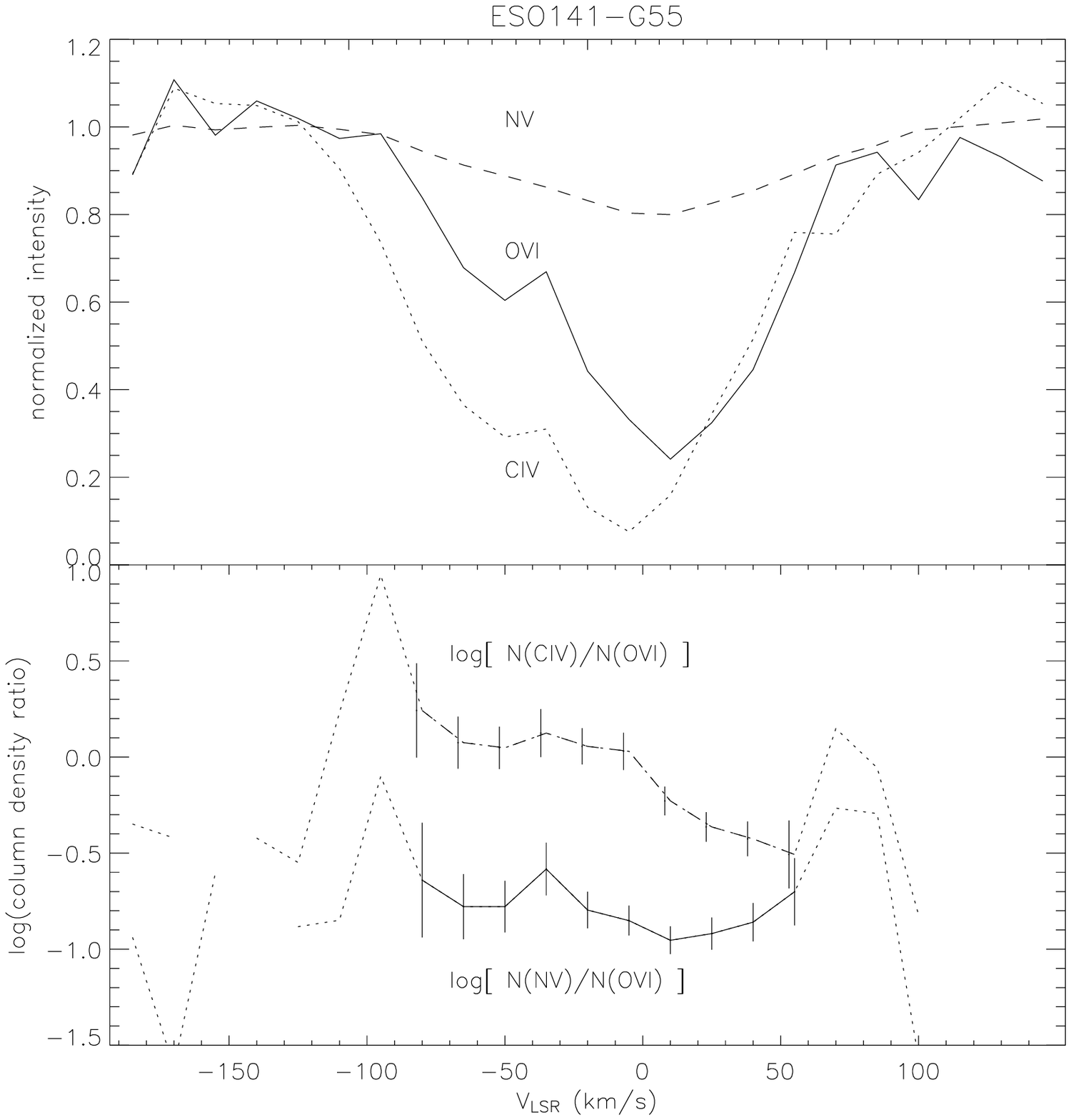}
\plotone{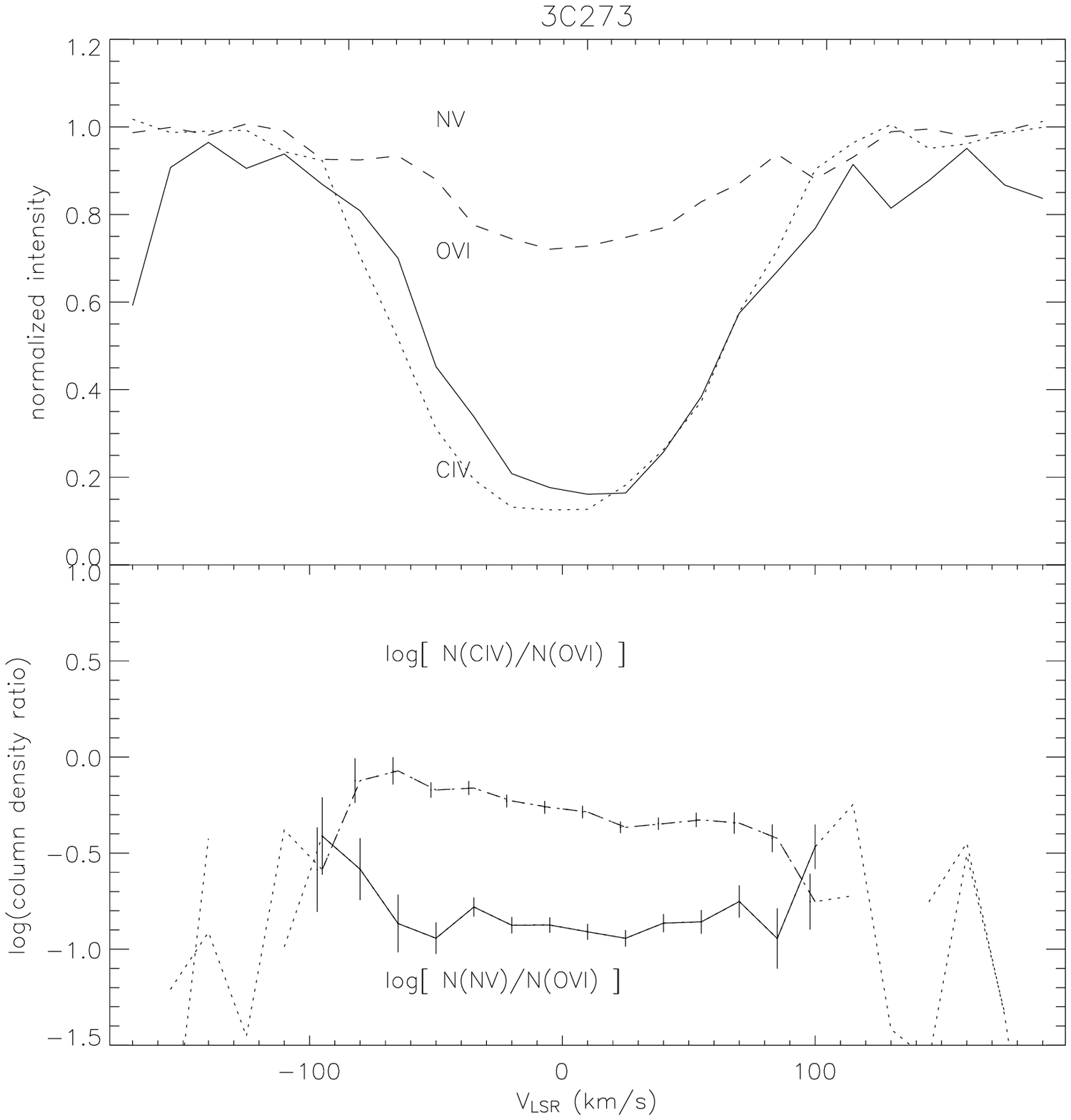}
\epsscale{1.0}
\caption{\label{sightlines}
\footnotesize
 Example sightlines.  The upper panel of each plot shows the
 normalized intensity of the three absorption lines, and the lower
 panel shows the log of the apparent column density ratio as a
 function of velocity (15 km~s\ts{-1} bins). The column density ratio
 is only used for each sightline over the velocity range for which the
 errors are sufficiently small, and where there is appreciable absorbing
 column (solid line).  Dashed lines indicate regions where the
 signal-to-noise (column density) is too low to use the line ratio.}
\end{figure}

Figure \ref{sightlines} shows two example sightlines, towards 3C273
and ESO141-G55.  The upper panel of each plot shows the normalized
intensity of the three absorption lines (\C\ was available for these
particular sightlines), and the lower panel shows the log of the
apparent column density ratio as a function of velocity. The column
density ratio is only used for each sightline over the velocity range
for which the errors are sufficiently small and there is appreciable
absorbing column. (Ratios are generally restricted to points with less
than 0.5 dex~(km~s\ts{-1})\ts{-1} error, which in practice limited
errors to $\lesssim$0.3 dex~(km~s\ts{-1})\ts{-1} in most sightlines
because the errors climb quickly in the line wings.)  Some of the
sightlines also include high velocity clouds, for which the column
density ratio can occasionally be measured.  As discussed below, these
objects are interesting because they may consist of old material
accreting onto the Galaxy for the first time, or possibly material
ejected to particularly high altitudes by a Galactic fountain.

\begin{figure}
\plotone{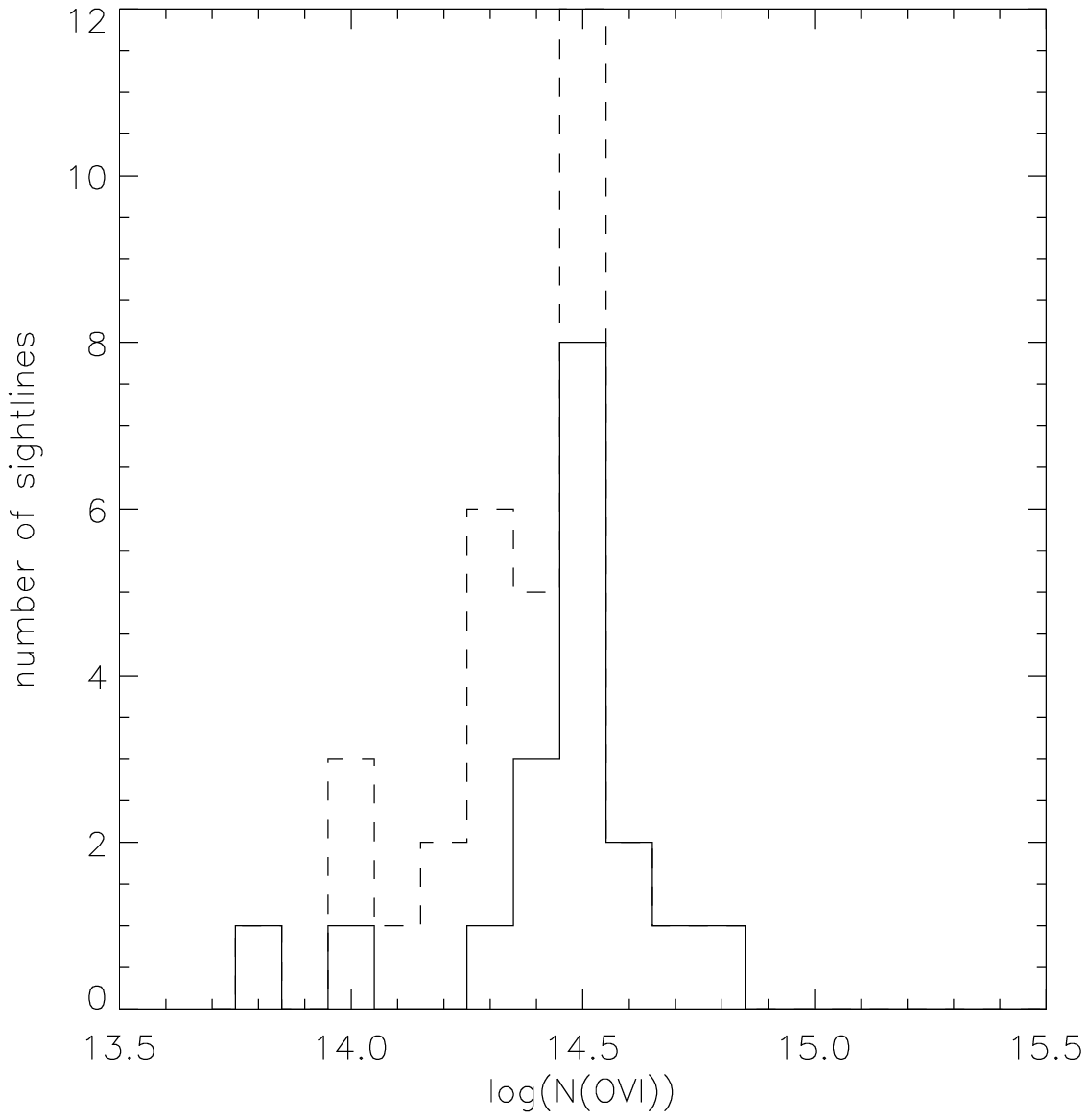}
\plotone{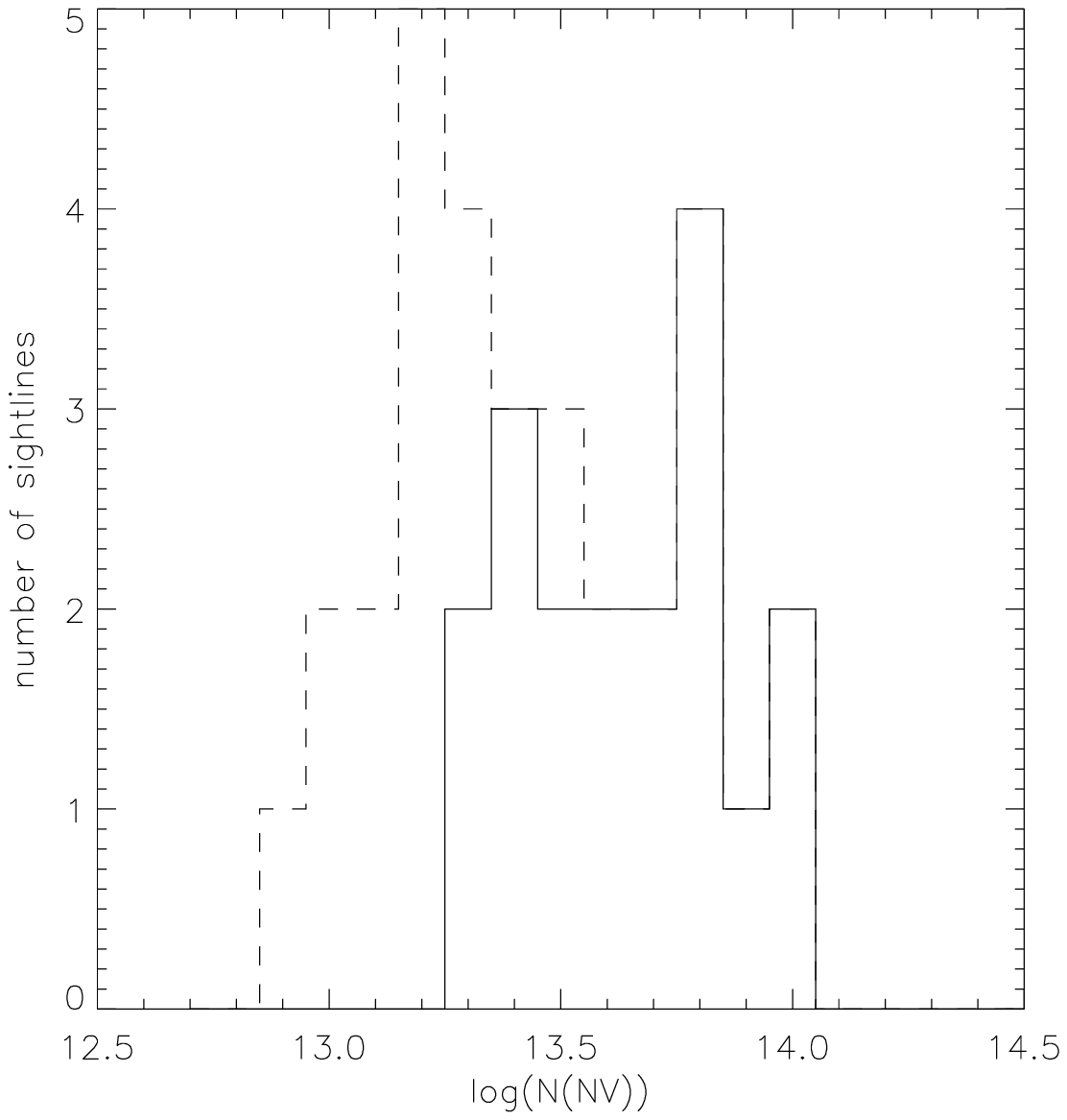}
\epsscale{1}
\caption{\label{histo}
\footnotesize
 Distributions of total column density of \N\ and \O\ along the 34
 sightlines for which measurements could be made (dotted line). 
 The distribution of column density for the sightlines with 
best measured N(\N) ($\delta$N(\N)$<$0.35 dex) 
 is shown as a solid line, and is naturally weighted towards the 
 higher column densities.
See text for properties.
 }
\end{figure}

Although the main focus of this study is the ratios of column
densities, brief discussion of the total line-of-sight absorption
measurements and how they relate to previous measurements is merited.
Figure \ref{histo} shows the distributions of total column densities
of \N\ and \O.  The medians of the distributions are
2.6\up{14}~cm\ts{-2} of \O\ and 2.8\up{13}~cm\ts{-2} of \N.
\citet{savage02} find a mean \O\ column density of 2.3\up{14}~cm\ts{-2}
in 91 sightlines including these, but omit the high velocity gas.
Theoretical models of evaporating gas predict $\sim$2\up{13}~cm\ts{-2}
of \O\ and $\sim$2\up{12}~cm\ts{-2} of \N\ per interface or cloud,
models of cooling gas predict a few \up{14}~cm\ts{-2} of \O\ and a few
\up{13}~cm\ts{-2} of \N, and finally turbulent mixing layer models
predict 10\ts{11}-10\ts{12}~cm\ts{-2} of \O\ and
10\ts{10}-10\ts{11}~cm\ts{-2} of \N.  (See Paper~I for a complete
description of the different models.)  The current observations cannot
distinguish between the models based on column density alone, although
the large number of turbulent mixing layers required is somewhat
worrisome.  As described in Paper~I and in the following section, the
trends of the column density ratios with velocity can be a more
powerful diagnostic of the physical production mechanism for Li-like
ions.

\begin{figure}
\plotone{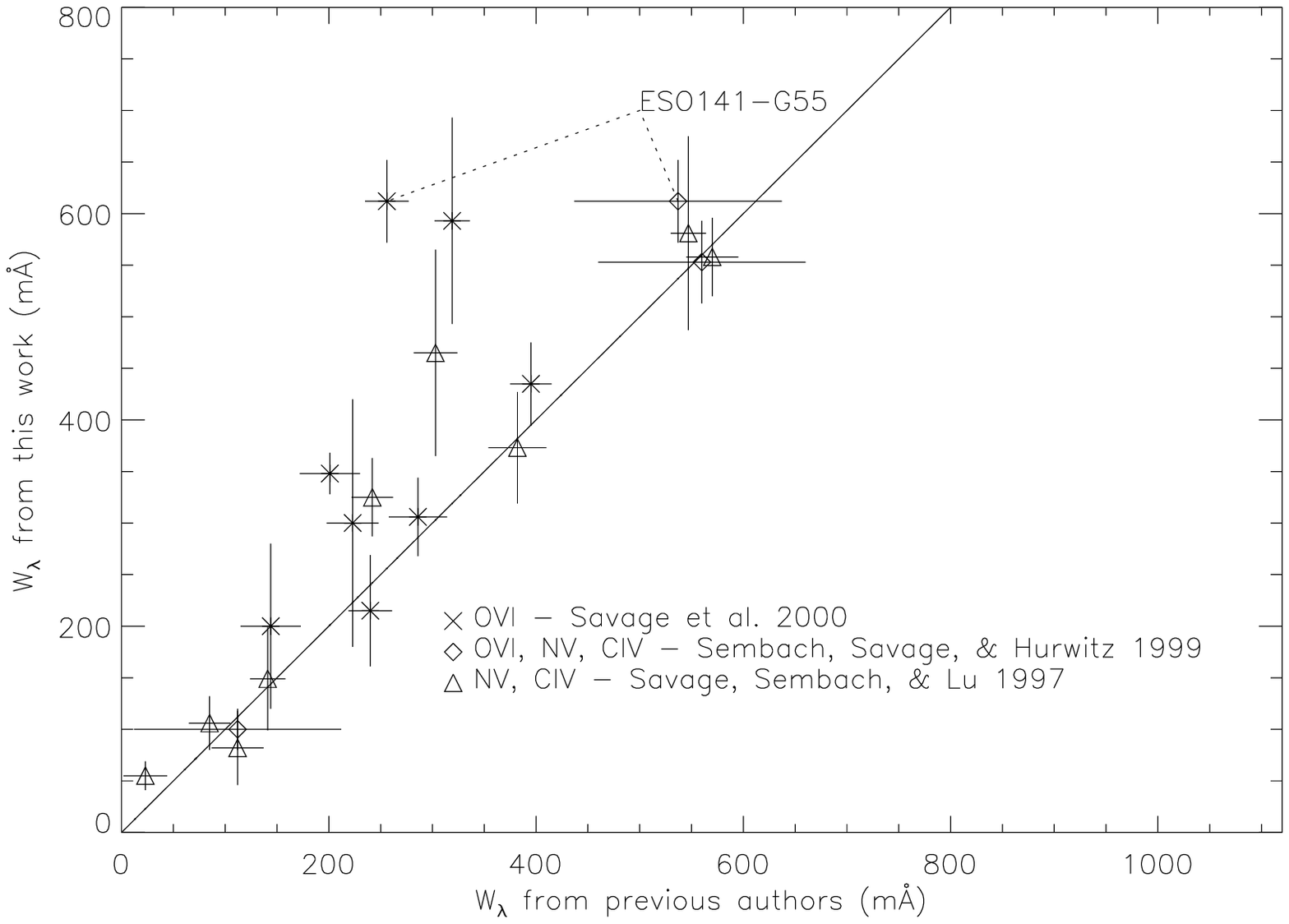}
\caption{\label{Wlam}
\footnotesize
 Comparison of equivalent widths for \C, \N, and \O\ measured in this
 and previous studies \citep{fuse2,SSH99,SSL97}.  Each point
 represents a single sightline.  The solid line shows a 1:1
 correspondence.  The effect of arbitrary velocity cutoffs is shown
 for ESO141-G55 (two different cutoffs from two different articles).
 Our analysis often gives higher equivalent widths because of these
 cutoffs. All \O\
 column densities also agree within errors with \cite{savage02} and
 \cite{wakker02}, taking into account velocity integration ranges.}
\end{figure}

Figure \ref{Wlam} compares the measured equivalent widths of some of
the sightlines in this work to previously measured values.  Each point
represents a single sightline, and measurements of \C, \N, and \O\ are
included in the plot.  There is a slight trend towards systematically
higher values in this study than previous work.  This results because
previous studies arbitrarily cut off absorption at some velocity
in an attempt to isolate Galactic halo gas from high velocity clouds
(HVCs).  Our study excluded high velocity gas only when that gas was
clearly separated in velocity.  The effect of arbitrary velocity
cutoffs is seen in two separate reportings of \O\ towards ESO141-G55
by other authors \citep{fuse2,SSH99}.  The higher equivalent width
value resulted from a less restrictive velocity cutoff, and that
higher value more nearly agrees with the value found in this work.

\subsection{Column Density Ratios}
\label{overalltrend}

\begin{figure}
\epsscale{0.79}
\plotone{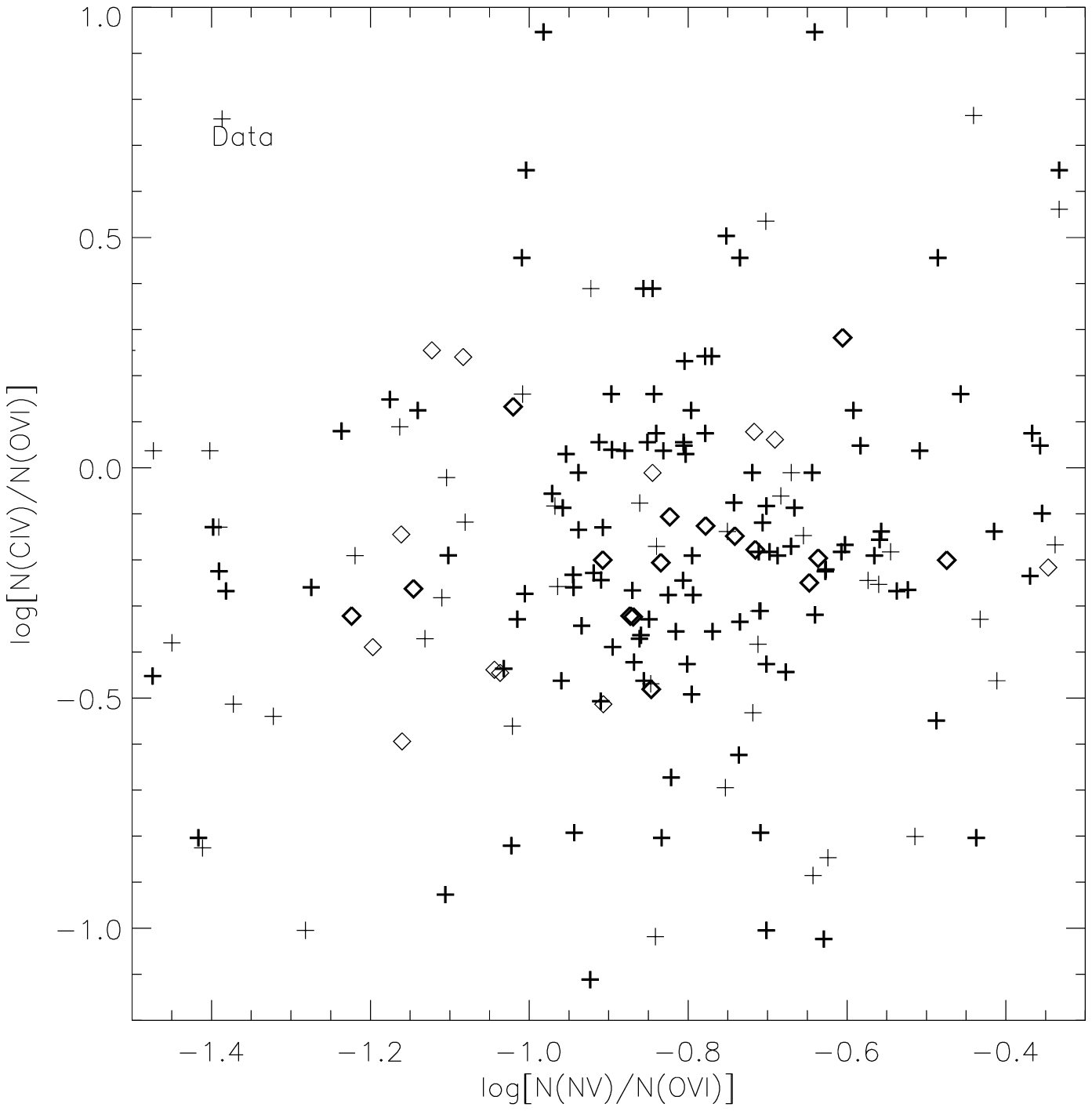}
\epsscale{0.85}
\plotone{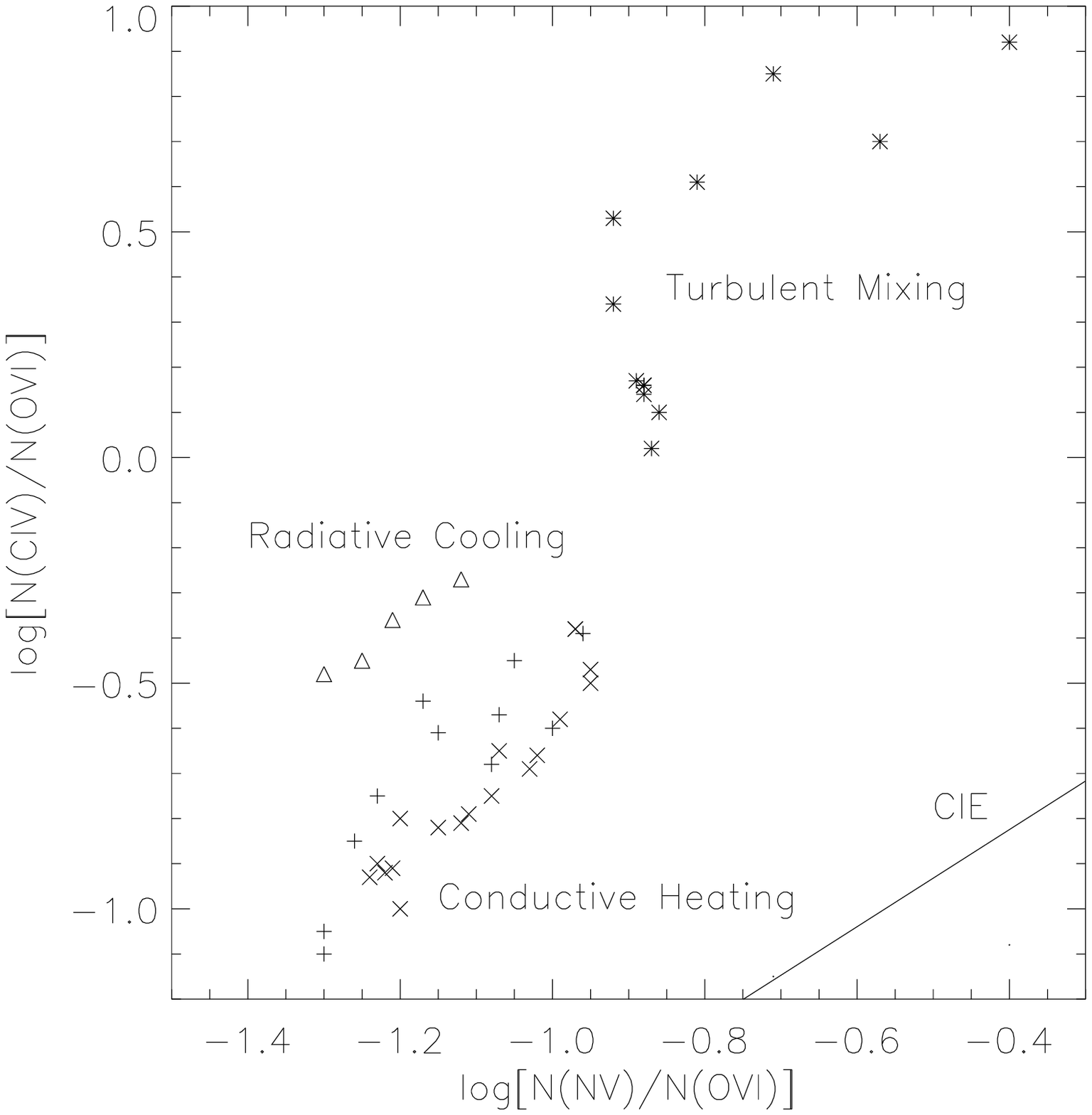}
\caption{\label{rat-color}
\epsscale{1.0}
\footnotesize
  First panel: Column density ratios for the sightlines along which all
  three ions were measured.  Each cross is a 15 km~s\ts{-1} velocity
  resolution element.  The dark diamonds are the velocity-integrated line
  ratios for each sightline.  Darker symbols are data for the sightlines 
  with highest signal-to-noise \N\ (error in total N(\N) $<$ 0.35 dex).
  Ratios should be compared with theoretical models presented in 
  Paper~I, repeated here as the second panel.  
The models are radiative cooling of
galactic fountain gas \citep[triangles,][]{shapiro93,benjamin93},
turbulent mixing layers \citep[stars,][]{mixing}, and conductive
heating and evaporation of spherical clouds \citet{bohringer87} and planar,
clouds \citet{borkowski90} (``x''s), and in cooling supernova remnant
shells \citep[pluses,][]{slavincox93,shelton98}.  Also shown are the
ion ratios for hot gas in collisional ionization equilibrium
\citep[CIE, solid line]{sutherland93}.
}
\end{figure}

\begin{figure}
\epsscale{0.8}
\plotone{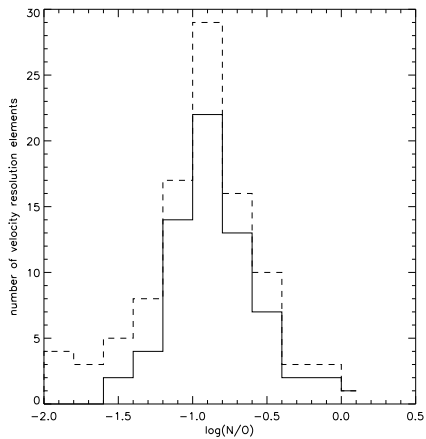}
\epsscale{1.0}
\caption{\label{rat-histo}
\footnotesize
 Distribution of velocity-resolved column density ratio
 log[N(\N)/N(\O)].  Each velocity resolution element is an instance of
 gas with a certain ion ratio, and this is a histogram of those
 instances.  The range of observed ion ratio covers the entire range
 in theoretical models (Paper~I).  Variation in the ion ratio with
 velocity is discussed in the text.  All sightlines are included in
 the dotted histogram, and only the highest signal-to-noise sightlines
 (error in total N(\N) $<$ 0.35 dex) are included in the solid
 histogram.  Note that the statistical properties of the \N/\O\
 distribution do not change with the inclusion of lower
 signal-to-noise sightlines.
}
\end{figure}

All of the sightlines for which all three ions are measurable
are combined in Figure \ref{rat-color} and compared to predictions
from several generic models.  The distribution
of velocity-resolved log[N(\N)/N(\O)] is shown in Figure
\ref{rat-histo}.  In Paper~I, it was shown how the ratios
of the column densities of Li-like ions could be used to distinguish
between various theoretical models of the production of these ions.
For example, models of thermally conducting, evaporating gas can be
distinguished from turbulent mixing layers.  Unfortunately, the data
span most of the range allowed by the models, and the predominance of
one type of model in the Galactic halo cannot be determined.  The
large thermal width of gas at several 10\ts{5}~K, compared to the
velocity separation of the different physical processes which almost
certainly exist along these lines of sight, precludes using Li-like
ion ratios to distinguish between models in the two-dimensional space
shown in Figures \ref{rat-color} and Paper~I.  It is important to note
that, in Figure~\ref{rat-histo} and in subsequent analysis, the
statistical properties of the \N/\O\ distribution do not change with
the inclusion or exclusion of lower signal-to-noise sightlines.

Table \ref{averages} lists the average column densities and column
density ratios for this sample of halo sightlines, as well as the
RMS dispersion of those quantities.  The total, or integrated, column
density along each line of sight is calculated, and the log of that
quantity is averaged over all sightlines.  The ratio of integrated
column density is taken for each sightline, and the log of that ratio
is also averaged over all sightlines.  Finally, the column density
ratio is calculated for each velocity resolution element along the
line of sight, and the log of that quantity is averaged, over all of
the velocity bins along all lines of sight.

\begin{deluxetable}{lcc}
\tablewidth{0cm}
\tablecaption{\label{averages} Average Halo Values}
\tablehead{
\colhead{ion or ratio} & 
\colhead{mean} & \colhead{RMS} 
}
\startdata
$\log$[N(\O)]~\tn{a}         & 14.54 & 0.19 \\
$\log$[N(\N)]                & 13.71 & 0.31 \\
$\log$[N(\C)]                & 14.33 & 0.17 \\
$\log$[N(\N)/N(\O)]~\tn{b}   & -0.83 & 0.28 \\
$\log$[N(\C)/N(\O)]          & -0.28 & 0.18 \\
$\log$[N(\N)/N(\O) (km~s\ts{-1})\ts{-1}]~\tn{c} & -0.76 & 0.64 \\
$\log$[N(\C)/N(\O) (km~s\ts{-1})\ts{-1}]        & -0.36 & 0.64 \\
\enddata
\tablenotetext{a}{Integrated column density along a sightline.}
\tablenotetext{b}{Ratio of integrated column density along a sightline.}
\tablenotetext{c}{Ratio of column density in each velocity resolution element along a sightline -- see text.}
\end{deluxetable}

Although the integrated column density ratio of a given theoretical
model is not as discerning a diagnostic as one might have hoped,
different physical scenarios should have different ionization states
as a function of velocity.  Offsets in velocity between the Li-like
ions, or gradients in the column density as a function of velocity,
could provide a more powerful diagnostic of the physical state of the
gas than the absolute value of the ion ratio, and this is modeled in
\S~\ref{interpretation} and Paper~I.  The velocity of the
absorbing gas can also be mapped onto Galactic altitude or radius,
assuming that the halo gas corotates smoothly on cylinders with the
disk.

Differences in velocity centroid and line width of the Li-like ions
have been marginally detected in previous work.  \citet{SS92} find an
increase in line width with increasing ionization stage, and their
data ($\langle v\rangle$ in their Table 6) reveal a small trend for
\N\ to be found at more negative velocities than \C.  Their data are
more confused than the sample studied here because of the choice of
sightlines.  They use some stars towards the Galactic center which may
not have the same gas kinematics as the outer disk
\citep[see][]{trippSS93}, and some stars in the Magellanic Clouds,
which may have their own hot gas kinematics (if those galaxies have
fountain flows or outflows, the gas kinematics are in fact most likely
to be opposite to those in our galaxy).  In summary, there is evidence
of differences in distribution between \C\ and \N, but the
interpretation is not clear.  The inner Galaxy sightlines of
\citet{trippSS93} are not inconsistent with \citet{SS92}, but the
trends are even less conclusive.
\citet{SSL97} also note that \N\ absorption is wider in velocity
extent than \C. Their data (their Table 4) show no particular
velocity offset between \C\ and \N.

\begin{figure}
\plotone{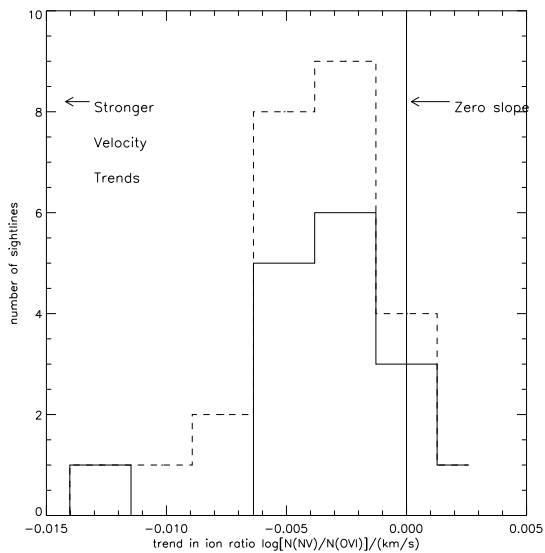}
\caption{\label{histo-slope}
\footnotesize
 Distribution of slopes fitted to the ion ratio log[N(\N)/N(\O)] per
 unit velocity along each sightline.  
 The mean slope is -0.0032$\pm$0.0022(r)$\pm$0.0014(sys)~dex~(km~s\ts{-1})\ts{-1}.
See text for discussion of random and systematic errors.
All sightlines are included in the
 dotted histogram, and only the highest signal-to-noise sightlines
 (error in total N(\N) $<$ 0.35 dex) are included in the solid
 histogram.  The offset exists in the high signal-to noise data at the
 same significance as in the full dataset.  }
\end{figure}

In this dataset, there is some indication of
a trend in the data towards higher mean ionization state
(lower \N/\O\ and lower \C/\O) in more redshifted gas.
Many individual sightlines appear to have a trend of decreasing
log[N(\N)/N(\O)] with velocity (see Figure \ref{sightlines}, lower
panels).  This suggests the following analysis: the dataset is
collapsed by making a linear fit to the ion ratio (in log space) as a
function of velocity along each sightline.  Clearly, there is no
reason for the log of the ion ratio to have a linear dependence on
velocity, but this allows a simple characterization of any general
slope or trend with velocity along a sightline.  Figure
\ref{histo-slope} is a histogram of the slopes fitted to the
log[N(\N)/N(\O)] ratio along each sightline, and the individual fits
are listed in Table~\ref{sightlines}.  There appears to be an excess
of sightlines with negative slopes, i.e. lower N(\N)/N(\O) at more
positive velocities.  The mean slope is
$-0.0032\pm0.0022$(r)$\pm$0.0014(sys)~dex~(km~s\ts{-1})\ts{-1}.
Random error (r) results from wavelength calibration,
continuum fitting, signal-to-noise of the data, and the goodness of
fitting a linear slope (which depends on the complexity of the given
line of sight and the specific physical structures observed in
absorption).  Systematic error (sys) is a generous estimate of the
remaining FUSE wavelength calibration problems and their impact on the
relative wavelength calibration between FUSE and HST.  Some of the
lowest signal-to-noise sightlines have been omitted from this
analysis, as indicated in the Appendix.

\begin{figure}
\plotone{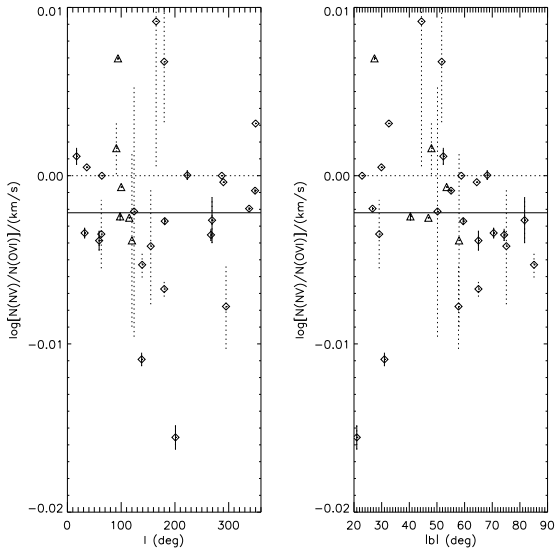}
\caption{\label{lbslopes}
\footnotesize
 Fitted slopes to the column density ratio log[N(\N)/N(\O)] as a
 function of Galactic longitude $l$ and latitude $|b|$.  Sightlines
 that pass through high velocity cloud Complex C are distinguished by
 triangles.  No systematic trends with Galactic location or with
 membership in Complex C are seen.  (None are particularly expected,
 see Paper I.)  The mean slope is marked with a solid line, and is
 negative, with lower N(\N)/N(\O) at more positive velocities. Only
 the best data (error in total N(\N) $<$ 0.35 dex) are included in that
 average, but including all the sightlines doesn't change the mean
 within error.}
\end{figure}

Figure \ref{lbslopes} shows the fitted slopes to the column density
ratio for each sightline as a function of Galactic longitude $l$ and
latitude $b$.  There are no particular trends, which could result from
selection effects or irregular sky coverage.  There is a dipole moment
in high-velocity cloud \ion H1 velocities, with more HVCs at negative
velocities in the northern 1\ts{st} and 2\ts{nd} quadrants, and more
at positive velocities in the southern 3\ts{rd} and 4\ts{th}
quadrants.  However, trends in the column density {\it ratio} with
velocity should not be directly affected by this dipole sky (see
\S~\ref{corotate} for more subtle effects), and no trends are seen
here.  Also singled out are sightlines that pass through HVC Complex
C.  Unlike the other major high-velocity clouds (the Magellanic Stream
in particular) detected in this data and excluded from the general
analysis, gas from Complex C cannot easily be discriminated by
velocity from gas over the outer Galactic disk. Complex C gas is
therefore included in our general analysis.  Its low metallicity
\citep[0.1--0.3 solar,][]{wakker99,gibson01,collins03}
argues that Complex C is a mixture of Galactic fountain gas and more
primordial infalling gas.  Thus, one might expect different
characteristics of the Complex C sightlines, but none are apparent.
Either the trends along those sightlines are not being dominated by
Complex C, or the trends do not form an appropriate diagnostic to
distinguish different conditions in Complex C. The Li-like
ionization structure in Complex C is apparently 
not significantly different from
the rest of the Galactic halo.

\subsection{Systematic and Instrumental Effects}

\begin{figure}
\epsscale{0.6}
\plotone{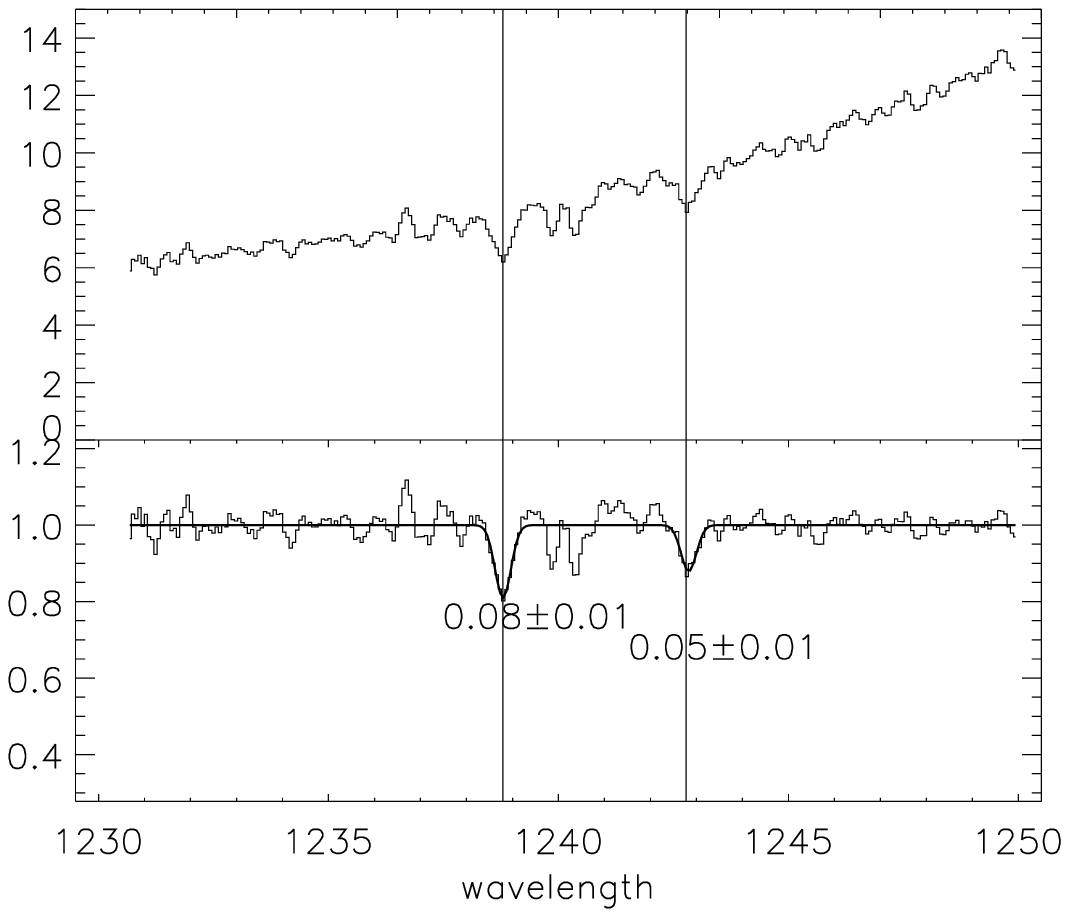}
\plotone{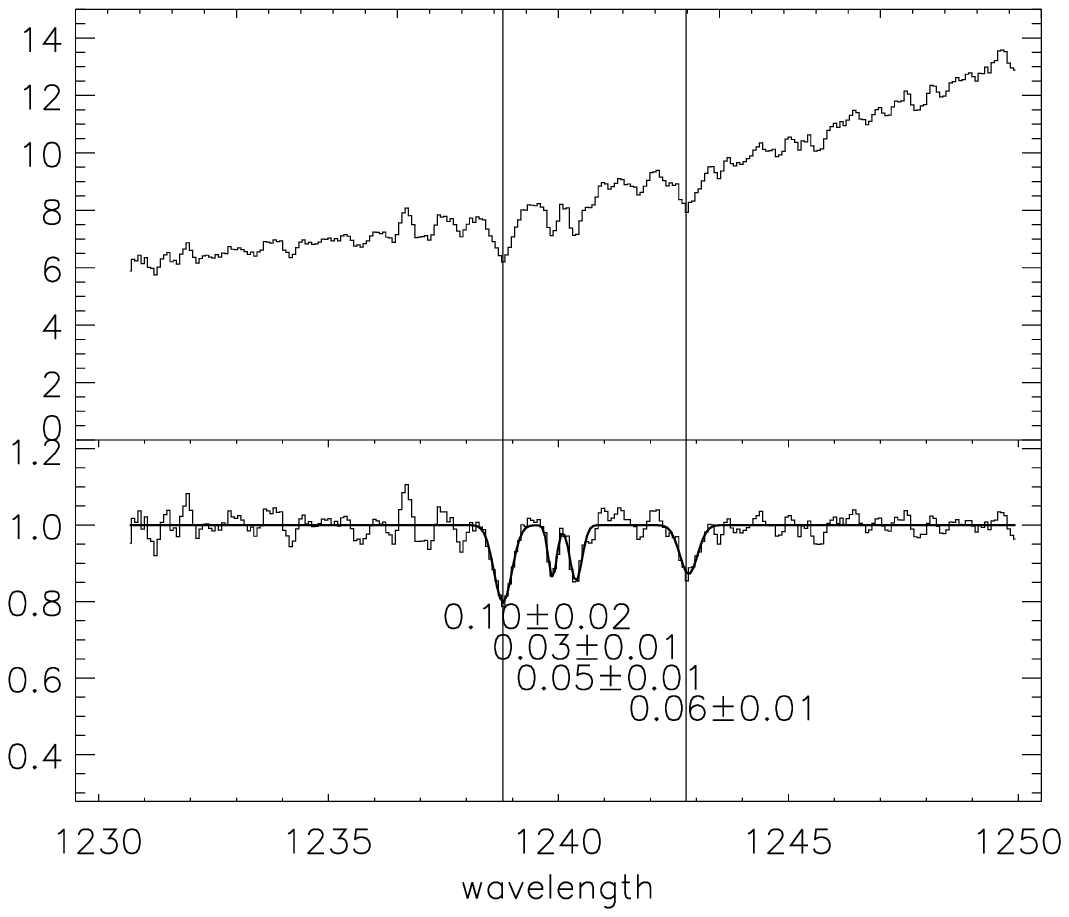}\\
\epsscale{0.7}
\plotone{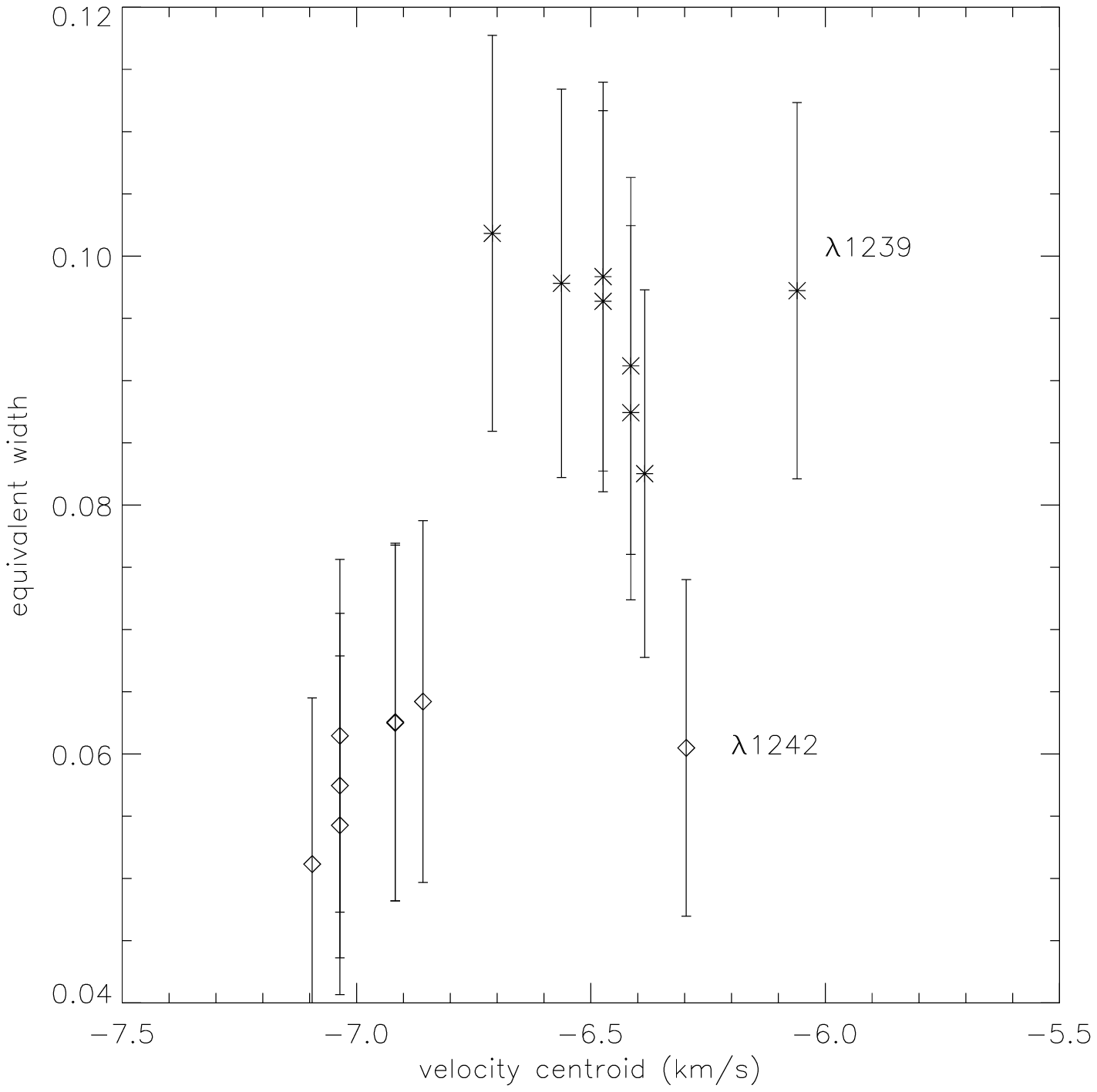}
\epsscale{1.0}
\caption{\label{continuum}
\footnotesize
Continuum fitting procedure.  The upper panels show the \N\ doublet
($\lambda\lambda$1238,1242) towards ESO141-G55, and two weak
\ion{Mg}{2} lines between the \N\ lines.  The equivalent widths (\AA)
of the \N\ lines are shown for the simultaneous fit of a 4\ts{th}-order
polynomial to the continuum and two or four Gaussian absorption
features.  The lower panel shows the fitted equivalent width and
central velocity for each member of the \N\ doublet, for various
combinations of continuum polynomial order and number of adjacent
absorption features.  The variation in the equivalent width is smaller
than the plotted error calculated from the fit.  The formal errors
for the central velocity are several times the horizontal size of this
plot and are not shown.  The variation in the equivalent width was
studied for each ion along each sightline. The continuum fit which
yielded values closest to the mean was used in subsequent analysis,
and the variation in equivalent width was incorporated into the quoted
error in total column density.}
\end{figure}

Continuum placement is often the dominant error in absorption-line
measurements, and it is especially important in the case of broad weak
lines such as those expected from interstellar \C, \N, and \O.  The
continuum in these data was fitted in a region $\sim$20~\AA\ wide
around the line of interest (the actual width was set by the
wavelength coverage of the observation and the desire to omit strong
emission features in the background source).  The continuum was fitted
as a fourth-order polynomial simultaneously with Gaussians for all
absorption lines in the region.  As demonstrated in Figure
\ref{continuum}, the inclusion or exclusion of neighboring absorption
features to the line of interest, especially weak features, was
explored and the variation which this caused in the equivalent width
of the line of interest was included in the experimental error.  We
also tried different order polynomials in the continuum fit and varied
the size of the fitting region.  Note that the Gaussian line-fitted
equivalent widths were only used to assess error due to continuum
fitting, and the final total column densities quoted below are
integrated across velocity.

Various other interstellar absorption lines can contaminate the lines
of interest, especially \O.  The \N\ and \C\ doublets should be clear
of other Galactic absorption, and for \N\ the doublets were carefully
examined for the presence of intergalactic Ly$\alpha$ or Ly$\beta$ absorption
(as regards Ly$\beta$, see in particular disussion of RXJ 1230.8+0115 in the Appendix).
Absorption from H$_{\rm 2}$ and \ion C2 frequently contaminates the
\O\ 1038~\AA\ line, so analysis was limited here to the 1032~\AA\
line.  Molecular hydrogen lines can also contaminate the 1032~\AA\
line, specifically Lyman~(6-0)P(3) 1031.19~\AA\ and Lyman~(6-0)R(4)
1032.35~\AA.  The strengths of these lines were measured using
isolated H$_{\rm2}$ lines of the same $J$~level elsewhere in the FUSE
spectrum, typically $\sim$10 lines for each rotational level for each
sightline.  The column density in that rotational level was
calculated.  The H$_{\rm 2}$ absorption is removed from the \O\
absorption where relevant, and the error bars of those points
increased accordingly.  The strength of the H$_{\rm 2}$ lines or their
upper limits are reported in Table \ref{targets}.  This process is
more precise for the P(3) line, as it is not a particularly strong
$J$=3 line compared to other measurable lines in the bandpass.  The
R(4) line is one of the stronger $J$=4 lines available in the FUSE
long-wavelength segments, so the error in the column measured from
weaker lines is larger.  The R(4) line is much weaker than the P(3)
line, however, and usually does not confuse the analysis.  It should
be noted that, in a few sightlines, more than one H$_{\rm 2}$
component was justified by inspection of other isolated lines in the
FUSE bandpass.
Incorrect subtraction of H$_{\rm 2}$ in this study is most likely to
err towards underestimating the \O\ column, because conservative
determinations were made in situations where the H$_{\rm 2}$
absorption is complex.  The presence of undersubtracted H$_{\rm 2}$ in
the stronger line, at a velocity of $-212$~km~s\ts{-1} relative to \O,
would decrease any trend of decreasing N(\N)/N(\O), so if there is
such a systematic error, the correction of that error would only
strengthen the trend.  

The \ion{Cl}{1}~$\lambda$1031.507 absorption
could also contaminate the \O\ line. The \ion{Cl}{1} column density was
measured along each line of sight using other isolated lines in the
FUSE bandpass.  Only upper limits were obtained for this
contamination, and no subtraction was performed.  Unresolved saturated
absorption components can be washed out in an absorption line, leading
to an underestimate of the absorbing column density.  This problem has
been discussed in detail by various authors including \citet{ss91} and
\citet{jenkins96}.  The absorption lines of \C\ and \O\ can suffer
from this effect, especially if these species are present in
overionized gas. (In such cases, the gas is cooler than the
temperatures at which these species dominate in collisional ionization
equilibrium, so that the thermal widths are smaller or comparable to
the FUSE resolution of $\sim$15~km~s\ts{-1}).  Absorption features of
\N\ are less susceptible, owing to the intrinsic weakness of the
\N$\lambda\lambda$1240 doublet.  As described in
\citet{ss91}, unresolved saturated absorption can be detected and
corrected for by comparing two absorption lines of different
oscillator strengths.  In the case of the three doublets in this
study, the apparent velocity-resolved column density can be compared
in the two lines of the doublet, whose strengths differ by a factor of two.

\begin{figure}
\plotone{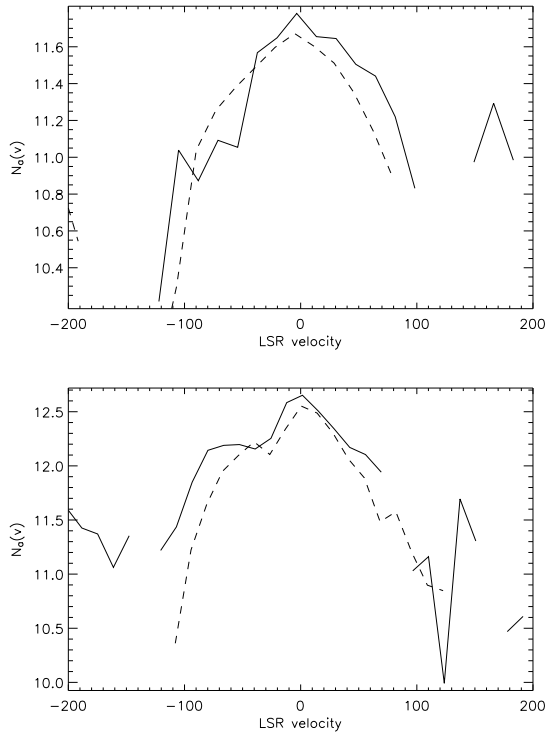}
\caption{\label{saturate}
\footnotesize
  Apparent column density as a function of velocity for \N\ and \C\
  towards ESO141-G55.  Column density calculated from the weaker line
  of each doublet (solid lines) is slightly greater than that
  calculated from the stronger line of the doublet (dashed lines),
  indicating unresolved saturated absorption components.  For the
  instrumental resolution and degree of saturation in this case, the
  methods of \citet{ss91} indicate that the correction to be added to
  the larger column density is about the same as the difference in
  apparent column density between the two lines, or $\lesssim$ 0.1
  dex (0.06$\pm$0.15 dex for \N\ and 0.1$\pm$0.1 dex for \C).}
\end{figure}

Figure \ref{saturate} shows the velocity-resolved apparent column
density $N_a(v)$ for the sightline that shows most clear evidence of
unresolved saturated absorption, ESO 141-G55.  The weaker line of the
doublet yields higher apparent column densities $N_a(v)$ than the
stronger line, by $\lesssim$0.1 dex (0.06$\pm$0.15 dex for \N\ and
0.1$\pm$0.1 dex for \C).  By the method of \citet{ss91}, the \C\ and
\N\ total column densities should be corrected by approximately 0.1
dex.  However, \O\ probably also suffers from unresolved saturated
components resulting in a similar correction. In almost all
sightlines, the \O\ 1038 line is confused by H$_{\rm 2}$ and
\ion C2 absorption, so the correction cannot be reliably made.  No
corrections for unresolved saturation are applied to the data in this
study, because the variation of column density ratio with velocity is
of primary interest, and because the correction for \O\ cannot be made
in most cases with only a single absorption line.  
We explore the effects of unresolved saturated components below, and
find it to be small.

\begin{figure}
\plotone{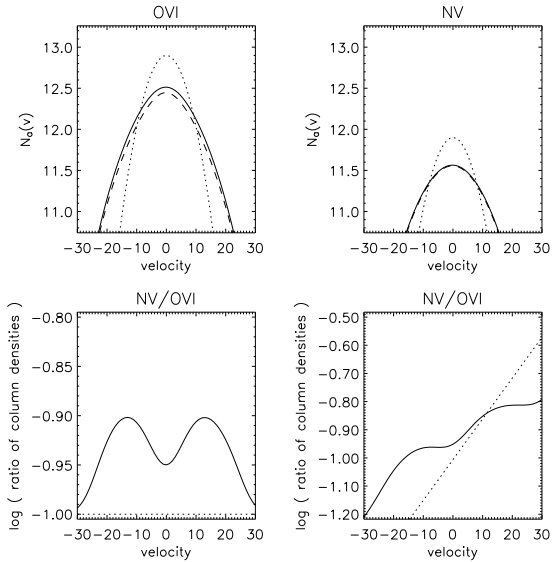}
\caption{\label{satmodel}
\footnotesize
  Modeled effect of unresolved lines shows that, in an unrealistically
  worst case, the velocity-resolved column density ratio can be
  miscalculated by 0.1 in the log.  In this model, similar to the
  analysis of \cite{ss91}, \O\ at the unrealistically low temperature
  of 25,000~K is observed by FUSE.  The apparent column density as a
  function of velocity $N_a(v)$ is plotted in the upper left panel, as
  calculated from the stronger (dashed) and weaker (solid) line of the
  doublet.  The true velocity-resolved column density $N(v)$ is also
  plotted with a dotted line.  The 0.1~dex difference in observed
  column density using the two members of the doublet is the largest
  saturation observed in \C\ data in this study, so this is also a
  worst case.  The upper right panel shows \N\ at the same temperature
  but 10 times lower concentration. The column density profile is
  blurred by the instrumental resolution but the total column would be
  accurately determined.  The lower left panel shows the log of the
  ratio of observed column densities, which differs from the true
  ratio of -1 by 0.1 dex.  More realistic temperatures and densities
  for the absorbing gas yield corrections $\lesssim$10\%, which is
  well within the errors calculated from other sources. 
  The lower right panel shows the true and observed log column density
  ratio for a situation in which the nitrogen central velocity is
  slightly offset from the oxygen.  The effect of instrumental
  resolution is to decrease the observed trend of ion ratio as a
  function of velocity.  }
\end{figure}

Figure
\ref{satmodel} shows the effect of unresolved absorption, in an
analysis similar to \cite{ss91}; the optical depth for \O\ at the
(probably unrealistically) lower limit temperature of 25,000~K is
moderately underestimated by being underresolved by FUSE.  At 10 times
lower concentration, \N\ is accurately measured.
In this worst-case scenario, the velocity-resolved column density can
be off by quite a bit (more than the total apparent column density,
because of the effects of instrumental smearing), but the effect is
symmetric across the line.  This effect cannot account for the large
scatter in observed \N/\O\ ratios (Figure~\ref{rat-color}).  
Most sightlines have lower quality data, and more realistic
temperatures for the absorbing gas result in column density ratio
corrections that, if they could be calculated for \O, would be within
the error bars.  \cite{wakker02} estimate that, for a set of
sightlines that includes this one, fewer than 20\% of the sightlines
should require saturation correction in \O, and the correction would
be less than 0.1~dex in those cases.

\begin{figure}
\plotone{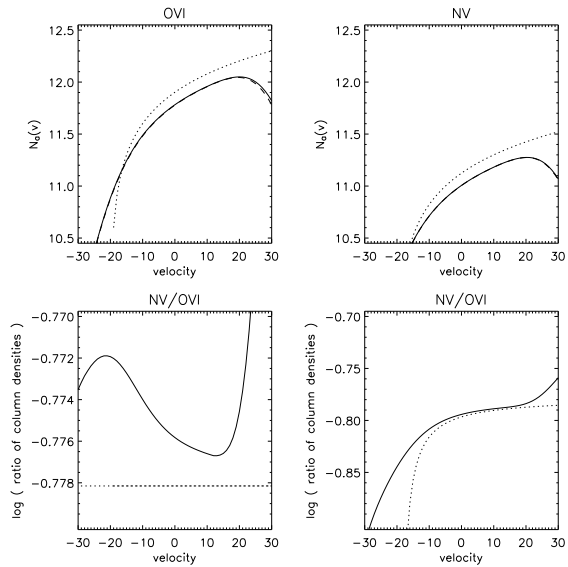}
\caption{\label{sat3}
\footnotesize
Modeled effect of saturation and instrumental smearing for an
intrinsically asymmetric line.  Panels are as Figure~\ref{satmodel},
for an absorption line with a sawtooth shape in $N(v)$, with typical
\O\ and \N\ column densities per km~s\ts{-1} (see
Table~\ref{averages}).  Smearing has a small effect on the measured
column, as seen in the top two panels, but the effect is similar
between \N\ and \O\ lines.
The apparent column density as a function of velocity $N_a(v)$ is
plotted as calculated from the stronger (dashed) and weaker (solid)
line of the doublet.  The true velocity-resolved column density $N(v)$
is also plotted with a dotted line.
The observed line ratio in the lower left panel shows a gradient not
present in the true column density distribution, but that gradient is
only $\sim$-0.0001~dex~(km~s\ts{-1})\ts{-1}, smaller than the gradient
measured in many of the sightlines in this study.  As for symmetric
lines, the lower right panel shows that asymmetric lines offset in
velocity also cause gradients in the line ratio, and that instrumental
smearing can decrease that gradient.  }
\end{figure}

In the context of velocity-resolved data, it is more interesting to
investigate the effect of saturation and instrumental smoothing on the
observed line ratio.  The fourth panel of Figure~\ref{satmodel} shows
that for symmetric line profiles, this effect will decrease any trend
of ion ratio as a function of velocity or wavelength.
If the line profile is not symmetric, then saturation and instrumental
smearing can affect the stronger \O\ line more than the \N\ line,
resulting in a small gradient in the observed ion ratio which is not
present in the true ratio.  Figure~\ref{sat3} illustrates this effect;
we model a line with a sawtooth shape in $N(v)$, with typical \O\ and
\N\ column densities per km~s\ts{-1} (see Table~\ref{averages}).
Smearing has a small effect on the measured column, as seen in the top
two panels, but the effect is similar between \N\ and \O\ lines.  The
observed line ratio in the lower left panel shows a gradient not
present in the true column density distribution, but that gradient is
only $\sim$-0.0001~dex~(km~s\ts{-1})\ts{-1}, smaller than the gradient
measured in many of the sightlines in this study.  Visual inspection
of the data shows that $\sim$1/3 of the sightlines have
an asymmetric \O\ line profile with more column at more positive
velocities, which in combination with instrumental smearing, could
produce a small negative slope in the observed \N/\O\ ratio.  As for
symmetric lines, the lower right panel shows that asymmetric lines
offset in velocity also cause gradients in the line ratio, and that
instrumental smearing can decrease that gradient.  In conclusion, the
reader should be aware of this effect in velocity-resolved data, but
we do not feel that it produces the entire trend seen in this
dataset.

A potentially critical systematic effect is the wavelength or velocity
calibration, since we combine data from different instruments.  The
FUSE wavelength calibration in older versions of the data pipeline in
particular was uncertain at the 5--10~km~s\ts{-1} level.  However, the
alignment of low-ionization interstellar absorption lines with \ion H1
21~cm emission can be used to correct the wavelength solution, at
least in the neighborhood of the Li-like ions.  The wavelength offset
for all data was corrected in the same way.  The wavelength offset was
measured in low-ionization, narrow interstellar absorption lines
expected to be present predominantly in neutral atomic gas:
\ion{Si}{2}, \ion O1, \ion C2, \ion{Ar}{1},
\ion{Fe}{2}, \ion N1, and \ion S2.
These were aligned with the LSR velocity of the neutral hydrogen
emission measured in the 21~cm line, either with the Leiden-Dwingeloo
survey \citep{LDS:proc} or the Parkes Multibeam survey \citep{parkes}.
We found the accuracy of this standard method of absolute wavelength
calibration to be $\lesssim$5--10~km~s\ts{-1}.
\cite{wakker02} made extensive tests of this calibration and found 
a similar accuracy. They also noted that version 1.8.7 of the FUSE
pipeline led to a 10~km~s\ts{-1} offset between \O\ 1032 and \O\ 1038.
Further investigation by scientists involved with FUSE pipeline
development (Kruk \& Sembach 2002, private communication) found that
the earlier software did indeed introduce a stretch in the wavelength
scale corresponding to 8~km~s\ts{-1} over the range spanned by the
low-ionization lines used (post-pipeline) to correct the wavelength.
(There was also an incorrect minus sign in the conversion from
spacecraft to LSR velocity.)  The correction as applied here did take
into account that wavelength ``stretch'' when data quality allowed it
to be measured.  Further, several sample sightlines were re-reduced
using version 2.0.5 of the FUSE pipeline, which has improved the
``stretch'' and wavelength calibration issues; no significantly
different offsets between \O\ and \N\ were discovered.  Although these
tests show no evidence of systematic wavelength calibration issues, we
estimate the maximum
systematic effect (due to remaining issues in the new pipeline or
perverseness of the data) to be 5~km~s\ts{-1}, which would introduce
an \ion N5/\ion O6 slope of 0.0014~dex~(km~s\ts{-1})\ts{-1}, less than
half of the mean effect seen in Figure~\ref{histo-slope}.  Thus, we
believe it unlikely that the entire trend seen here is due to
systematic problems in the FUSE calibration process, though we can
only rule this out with 80\% confidence. (There is a 20\% chance that
so many of our ion slopes would be negative if the true mean slope in
the Galactic halo was zero.)

\section{Implications for the Galactic Distribution of Hot Gas}
\label{interpretation}

The Li-like ions \C, \N, and \O\ are produced in nonequilibrium
physical situations, such as shocks, conductive interfaces and rapidly
cooling gas (Paper~I).  The different ions are generally
produced at different flow velocities in the structure, so a viewer
looking through such a structure will see trends in the ion column
density ratios as a function of line-of-sight velocity.  The
ratio-velocity signature in all possibly relevant physical situations
must be considered to properly interpret trends in the data. 

Trends in the column density ratios, and those ratios as a function 
of line-of-sight velocity, could also reflect differences in the 
spatial distributions of the different ions in the Galactic halo. 
In the following sections we interpret our dataset in terms of 
the large-scale distribution of hot gas in the halo.  We attempt 
(unsuccessfully) to model the observed trends in ion ratio with 
such a purely {\it kinematic} model, as opposed to the more physical 
models (shocks, conductive interfaces, etc) discussed in Paper~I.

\subsection{Corotating Smooth Halo}
\label{corotate}

As a reasonable starting point, one can imagine a smooth halo of
ionized gas distributed in an exponential distribution in Galactic
altitude, corotating on cylinders with the disk gas, without
postulating a physical mechanism for producing this distribution.  A
smooth density distribution requires fewer assumptions than a patchy
or clumpy one.  There is observational evidence of such patchiness
(for example the scatter in total column density and column density
ratio found in this study, and the extensive investigation of
\citet{howk02} showing large variations in \O\ towards the Magellanic
Clouds), but a smoothly distributed model may capture the averaged
global characteristics of the gas.  As for the second assumption, a
rotation rate independent of Galactic altitude is the simplest
starting point, although observational evidence for such corotation is
ambiguous.  For example, \citet{SSL97} argue that \C\ corotates with
the disk up to Galactic altitudes $|z| \sim$ 5~kpc, based on the lack
of strong secondary absorption peaks.  This is a well-known phenomenon
in which strong velocity gradients along the line of sight can create
discrete features in an absorption spectrum. These look like density
enhancements, even if the true space density of the gas is smooth or
constant.  It it not clear that the lack of secondary features can be
used in this manner to argue for corotating gas, or at least a small
velocity gradient, if the halo gas distribution is extremely patchy.
It is notable that there is significant \C\ absorption visible in
their figures outside of the velocity range allowed by corotation, and
that there is somewhat more of this gas at negative than positive
velocities.  (The authors also note that 7/9 sightlines show negative
\C\ velocities, with similar results for \N).  Significant
forbidden-velocity gas was also detected near the Galactic center by
\citet{trippSS93}.  Even a fairly smooth halo in a rising fountain
flow may not corotate with the disk; \citet{bregman80} notes that the
rotation speed decreases with altitude, mostly because the radial
gradient in the Galactic gravitational potential decreases with
altitude.

Despite the probable incompleteness of this model, it is instructive
both to attempt to determine the scale height of the hot gas,
and to model the ion-ratio signature that such a structure would have
in velocity space.

\subsection{Scale Heights}
\label{height}

\begin{figure}
\plotone{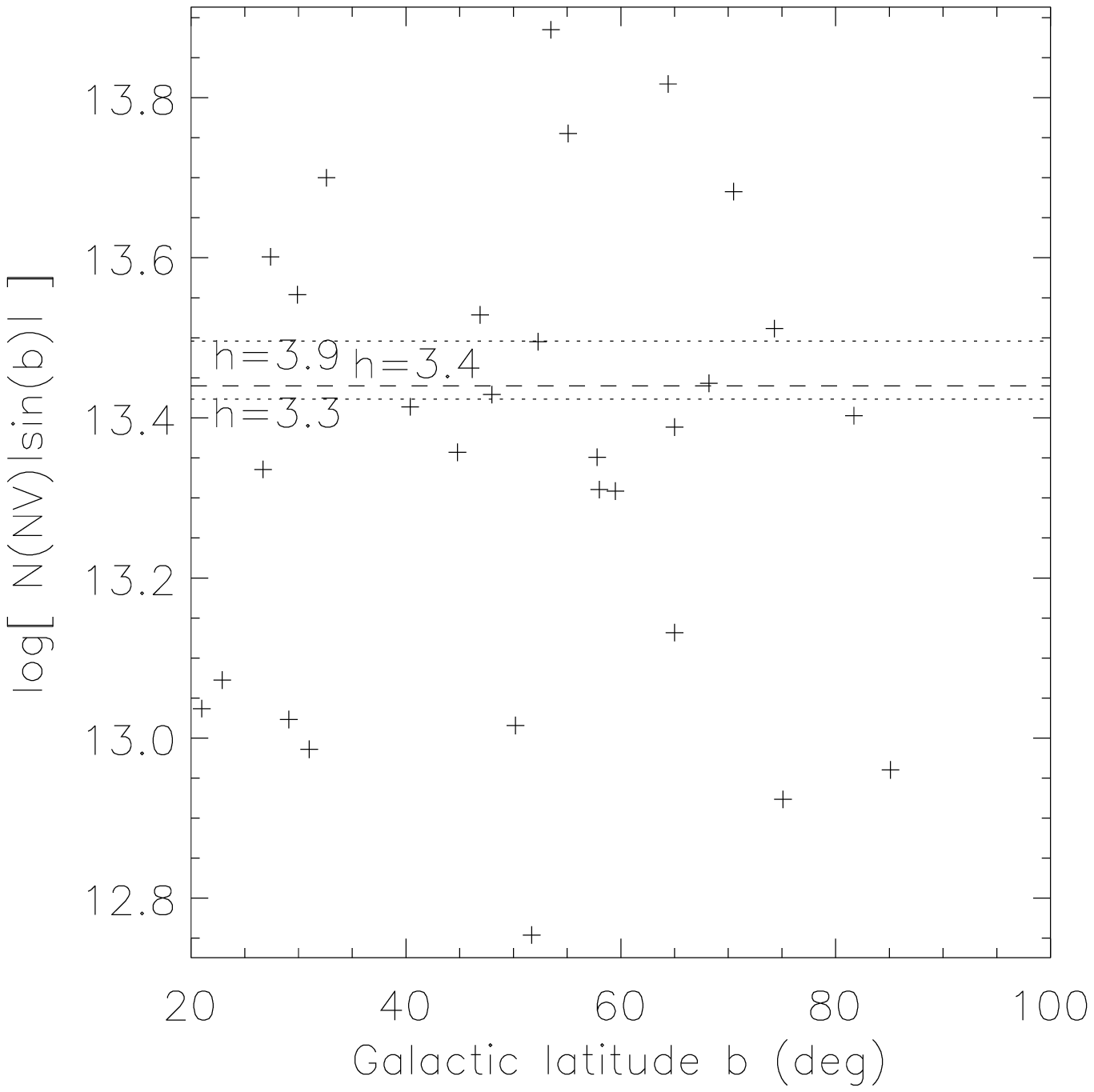}
\plotone{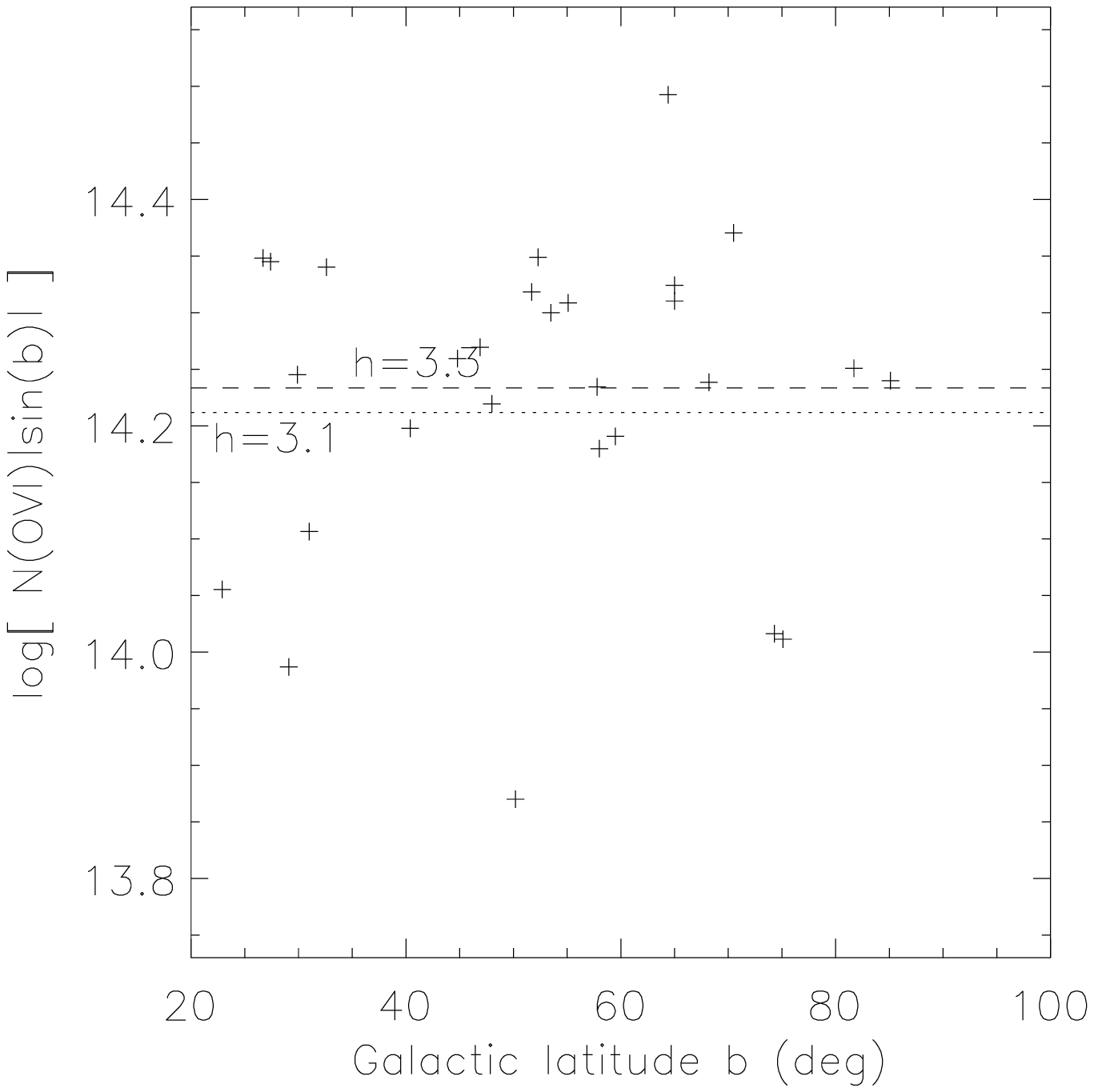}
\caption{\label{scaleht}
\footnotesize
 Scale heights in kpc for \N\ and \O\ determined from $N|\sin(b)|$.
 The value determined here is marked with a dashed line, and the
 values determined previously are dotted \citep{SSL97,fuse2}.  For \N,
 \citet{SSL97} determined a higher (lower) scale height depending on
 whether some of their upper limits were treated as measurements (or
 not).  The value found here falls between their values. Only the
 highest signal-to-noise \N\ data (error in total N(\N) $<$ 0.35~dex)
 was used to calculate the \N\ scale height (although including all
 sightlines doesn't change the answer to within error).}
\end{figure}

The scale height of the halo gas can be determined by fitting
$N|\sin(b)|$ versus Galactic altitude $z$ \citep[][for \N, \C, and
\ion {Si}{4}]{SSL97}.  One obtains a scale height and midplane density
from the fit.  Naturally, when only extragalactic targets are
observed, one takes the mean $N|\sin(b)|$ and divides by a given
midplane density to obtain the scale height, instead of obtaining the
midplane density directly from the data.  Figure \ref{scaleht} shows
$N|\sin(b)|$ for \N\ and \O\, and the derived scale heights using
midplane densities of $n_0$=1.7\up{-8}~cm\ts{-3} \citep{jenkinsparis}
for \O\ and $n_0$=2.6\up{-9}~cm\ts{-3} \citep{SSL97} for \N.  The
scale heights agree with previous work, confirming that the vertical
extent of gas increases with decreasing ionization stage, from \O\ to
\N\ to \C\ to \ion{Si}{4}.  Such trends were predicted by
\citet{shullslavin94}.  The large variations in $N|\sin(b)|$ result
from the extreme patchiness of the gas in even large scales.
\citet{SSL97} used a patchiness parameter in their fits to the scale
height, and found that almost all of the scatter is from intrinsic
patchiness rather than observational random error.  This is further
confirmed by \citet{savageparis}, who saw a hint that the \O\
distribution cannot be fitted by a plane-parallel structure (or
that there is large variation above different parts of the Galactic
disk), and by \cite{savage02}, in which patchiness is the clear cause
of the data being consistent with scale heights 2--5 kpc.  The extreme
patchiness of interstellar hot gas is also confirmed by measurement of
\O\ emission -- the ratio of emission (proportional to $n^2$) to
absorption (proportional to $n$) gives the space density $n$, and
shows that the gas is contained in small, dense structures, not a
diffuse smooth distribution \citep{shelton01,dixon01}.  This is
expected since this gas is commonly produced in thin interfaces or
highly disturbed cooling gas.

\subsection{Deconvolution of observations}

If one assumes that the hot halo gas corotates on cylinders with the
disk, line-of-sight velocity can be mapped onto Galactic altitude $z$
and Galactocentric radius $R_G$.  
This mapping does not reveal any trend of ion column density ratio 
with $z$ in this dataset.  
This does not rule out a
change in ionization as a function of Galactic altitude. It merely
demonstrates that a simple mapping based on smoothly distributed gas
corotating on cylinders above the disk is of limited use at high
latitudes.  Similarly, there is no clear trend in ion column density
ratio as a function of implied Galactocentric radius $R_G$.  This is
even less surprising, as the sample here consists of mostly
high-latitude sightlines.  It is known that there is a gradient in
Galactic metallicity, with a factor of 10 drop from Galactic center to
the outer Galaxy, resulting from the gradient in star formation
activity.  There are suggestions, but no incontrovertible evidence,
that the gradient might be steeper for nitrogen \citep[$-0.09\pm0.01$
dex~kpc\ts{-1},][]{rolleston00} or carbon \citep{hou01} than for
oxygen \citep[$-0.067\pm0.008$ dex~kpc\ts{-1},][]{rolleston00}.  The
dataset presented here may show hints of decreasing N(\C)/N(\O) at
larger $R_G$, consistent with a steeper gradient of carbon than
oxygen, but the trend is not statistically significant. (The mean
N(\C)/N(\O) decreases by less than 0.2 dex from $R_G$ of 5 to 100 kpc
while the variation along and between lines of sight is an order of
magnitude).  Even if this trend exists, it could indicate a change in
ionization and not metallicity as a function of $R_G$.  In this
dataset, it is probably an artifact of trying to extract information
from high-latitude sightlines with weak assumptions about corotation.

\subsection{Modeled velocity-ionization signature}

\begin{figure}
\plotone{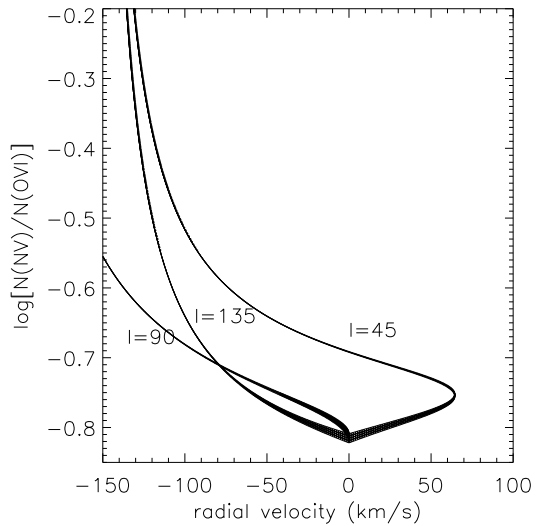}
\plotone{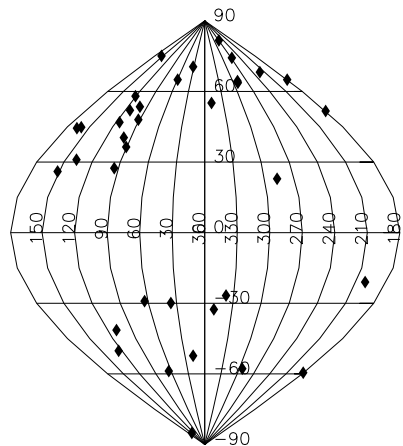}
\caption{\label{smooth}
\footnotesize
 Signature of a smoothly corotating halo with different scale heights
 for \N\ and \O.  The left panel shows log[N(\N)/N(\O)] as a function
 of radial velocity for sightlines at $b$=40\dg\ and various Galactic
 longitudes in the 1\ts{st} and 2\ts{nd} quadrants, for a smooth,
 corotating halo with a larger scale height of \N\ than \O.  The trend is
 reversed in velocity for sightlines in the 3\ts{rd} and 4\ts{th}
 quadrants, and reversed in the log of the column density ratio for
 southern sightlines.  The right panel shows the distribution of
 sightlines in this study.  }
\end{figure}

Mapping the data to Galactic altitude using corotation on cylinders
is questionable and does not produce illuminating plots. It is
more instructive to model what the velocity-ionization signature of a
smoothly corotating halo would be, to better understand what observations
could show. Figure \ref{smooth} shows the modeled trend in
log[N(\N)/N(\O)] as a function of line-of-sight velocity for
sightlines at $b$=+40\dg\ through a smooth, corotating halo in which
the scale height of \N\ is larger than \O\
\citep[\S~\ref{height}, ][]{fuse2,SSL97}.  The thickness of the line
scales with the density of \N\ and \O\ at that point along the line of
sight.  This figure also shows the distribution of sightlines in this
study.  There is an excess of sightlines in the northern sky near
$l\sim$120\dg, and this direction would display decreasing N(\N)/N(\O)
at more positive velocities in such a smooth halo.  It should be
recalled that these sightlines also pass through HVC Complex C, and
might be expected to have special characteristics for that
reason. However, no difference was detected in the column density
ratio along these sightlines compared to the general sample (Figure
\ref{lbslopes}).

\begin{figure}
\plotone{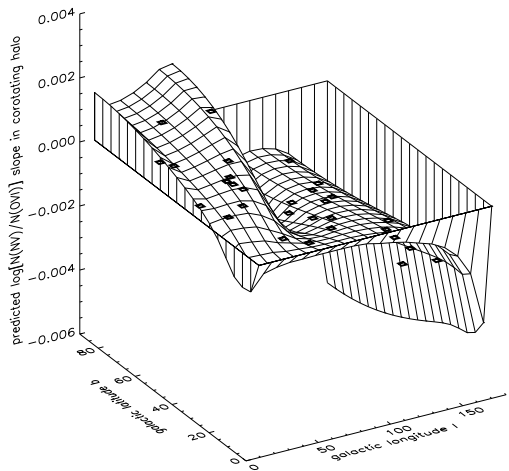}
\caption{\label{smoothcomp}
\footnotesize
Predicted slopes in the log of the column density ratio N(\N)/N(\O),
in a model in which the halo is corotating on cylinders with the disk, and
there ions follow smooth exponential distributions in Galactic altitude $z$
(different scale heights for each ion).
The predicted ion slopes at the Galactic positions of the observed sightlines
are marked.
}
\end{figure}

The velocity resolved ion ratio signature of a smoothly corotating
halo is further illustrated in Figure~\ref{smoothcomp}, in which the
predicted trend is shown as a function of Galactic longitude $l$ and
latitude $b$.  The predicted slope is symmetric in $b$ and
antisymmetric in $l$, so only $b>0$ and $0<l<180$ is shown.  Also shown are
the Galactic locations of the observed sightlines and their predicted
slopes. There is no correlation between the observed and predicted slopes.
\cite{sembach02} explore the possibility of fitting high velocity \O\
with a corotating halo model and find that the data do not support
(but also do not rule out) such an arrangement.

\section{Conclusion and Summary}

We consider the diagnostic power of velocity-resolved column density
ratios in understanding the Galactic halo.  Column density ratios of
Li-like ions in the Galaxy are useful to diagnose the physical
formation mechanism of the gas and to study the interstellar gas
cycle, and a survey of these ions can reveal general trends.  We find,
in a survey of sightlines observed with FUSE and HST, that the
distribution of N(\N) and N(\O) in the halo does not appear to favor a
dominant physical production mechanism.  We find a
possible weak trend of decreasing N(\N)/N(\O) at more positive
velocities
($-$0.0032$\pm$0.0022(r)$\pm$0.0014(sys)~dex~(km~s\ts{-1})\ts{-1},
Figure \ref{histo-slope}).  The weakness of this trend also argues 
against halo sightlines dominated by a single structure, e.g. the 
Local Bubble interface.

In Paper~I we presented models of interfaces and cooling
nonequilibrium gas, and the velocity-resolved N(\N)/N(\O) signatures
of each.  Here, we consider the model of smoothly distributed gas
corotating on cylinders with the Galactic disk, and implications for
the scale heights of \N\ and \O.
Observable velocity-ionization trends are weak, because even very
strong trends are washed out by the large thermal width of the gas at
different parts of the flow.  Additional confusion results because the
long sightlines almost definitely pass through multiple
structures.  In fact, the dispersion of N(\N)/N(\O), both
integrated and velocity-resolved, clearly indicates that no single
production scenario known to date can completely explain the Galactic
halo.  To truly understand the physical production of Li-like ions in
the halo, one needs to analyze gas in localized areas
of physical space, rather than velocity space.  Absorption
spectroscopy towards many halo stars with close angular separation and
different distances could help to isolate gas at a specific altitude.
Similarly, the gas above known superbubble shells or chimneys could be
isolated.  These observations have a greater chance of distinguishing
between models of hot gas production than observations along long
lines of sight.

\acknowledgments

We thank S. Penton, E. Wilkinson, J. Green, B. Savage, and K. Sembach
for useful discussions.  We thank the referee for a thorough reading
of the manuscript.  R.I. was partially supported during this
investigation by an NSF Graduate Student Fellowship to the University
of Colorado.  J.M.S. acknowledges support from NASA grant NAG5-7262
and NSF grant AST02-06042.  This work was based in part on data
obtained for the Guaranteed Time Team team by the NASA-CNES-CSA FUSE
mission operated by the Johns Hopkins University.  Financial support
to U.S. participants has been provided by NASA contract NAS5-32985.
We thank the members of the FUSE instrument, science, and operations
teams for their dedicated efforts.  We also thank the FUSE science
team and the principal investigators of the STIS and GHRS observations
for their permission to use these data, in some cases before public
release.

\clearpage

\appendix

\section{Notes on individual sightlines}
\label{appendix}

Some of the sightlines have previously published analyses, and some
pass through Galactic structures or high velocity cloud structures of
particular interest (see Figure \ref{smooth} for a plot of the
sightlines in Galactic coordinates).  The sightlines in this study
were each examined in the light of previous analyses to determine if
they had special characteristics warranting exclusion from a general
study, or if there were any discrepancies with previous
measurements. No instances of either were found.  Figures 24--31
show selected examples of the velocity-resolved column densities.  

Sightlines without notable Galactic interstellar features or data
characteristics are not discussed here.  One of the most obvious
differences between sightlines is whether they pass through Galactic
radio loops, which are probably superbubble shells filled with hot
gas.  Although these sightlines have typically higher total column
densities of Li-like ions than other sightlines, it is logical to keep
them in the sample for this study, because the tops of superbubbles
and the hot gas and shells that comprise them make up the 
Galactic fountain flow.  Sightlines that pass through radio loops
might preferentially sample rising material from the fragmented tops
of superbubbles, and other sightlines might preferentially sample more
evolved cooling gas returning to the disk, but our
understanding of the Galactic fountain is not precise enough to
justify any culling of our sample of halo sightlines.

{\bf 3C273:\ } The sightline towards the bright quasar 3C273 was
described in some detail by \citet{savage93}.  
It passes through radio continuum Loops I and IV \citep{berk71} and
the North Polar Spur \citep{heiles80,snowden95}.  Galactic radio loops
in general, and Loops I and IV in particular \citep{iwan80}, are
believed to be supernova remnants or superbubbles filled with hot gas,
so this sightline is expected to be strongly influenced by the
interface between that hot gas and the denser shell around it.
\citet{savage93} note that the high-ionization lines have average
velocities $\sim$10~km~s\ts{-1} more negative than weaker-ionization
lines, which they attribute to infalling hot gas.  From analysis of 
low-ionization lines, \citet{3c273} find elemental depletions typical
of warm diffuse halo clouds, consistent with a partial destruction of
grains (stripped mantles with cores remaining).  They also note that
\C, \N, and \O\ have similar absorption profiles except for a high
positive-velocity wing of \O, which they attribute to hot gas being
expelled from the Galaxy.

Despite the presence of the radio loops, there was no compelling
reason to exclude the 3C273 sightline from the general analysis of
this study.  (In particular, the proposals of other authors that this
sightline is involved with hot gas flowing in or out argues for its
inclusion in this study of halo gas dynamics.)  The resolved \N/\O\
and \C/\O\ ratios found here agree well with the integrated ratios of
\citet{3c273}, as does the integrated \O\ column density.  The
integrated \N\ column density agrees well with that of \citet{penton1}.

{\bf 3C351:\ } The sightline to this quasar passes through HVC Complex
C, the edge of Galactic radio Loop III, and through the outer warp of
the Galaxy \citep{savage93b}. Absorption associated with the outer
warp or with Complex C \citep[these two structures may join or be
related, see][]{wakker91} is seen at similar velocities as in
H1821+643, but the FUSE data are of insufficient quality for detailed
analysis.  This sightline is therefore excluded from the sample used
to search for general trends.

{\bf ESO141-G55:\ } The sightline towards ESO141-G55 passes near the
inner Galaxy and through the edge of radio Loop II.  \citet{SSH99} and
\citet{shull2000} note the presence of fairly cold molecular gas,
suggesting denser intervening material than other sightlines of
similar Galactic latitude, and possibly related to some enhancement of
Li-like ions relative to nearby lines of sight.  The ion ratios found
here are consistent with their integrated values, and no striking high
velocity absorption is noted.

{\bf Fairall 9:\ } The Fairall 9 sightline passes through the
Magellanic Stream, which accounts for positive velocity absorption in
the Li-like ions.  The apparently higher column densities reported
here compared to \citet{SSL97} are the result of not distinguishing
between the Galactic and Magellanic Stream components.  The numbers
found here for just the disk component agree with theirs.  The two
components are separated in velocity for \C, and possibly for \O, but
are difficult to distinguish in \N ~(indeed the \N\ detection is very
weak, and only the Galactic absorption was used in analysis of ion
ratios).  The large linewidths do not allow separation of different MS
high velocity clouds (HVCs) seen at 160 and 200~km~s\ts{-1} in 21~cm
emission \citep{morras83}.
The column density ratio log[N(\N)/N(\O)] has a lower limit of $-$1.7
in the high-velocity MS absorption, and log[N(\C)/N(\O)] ranges from
$-$0.5 to $-$0.9.  These values are lower than typical for gas
believed to be part of the Galactic halo and probably indicate a
higher ionization state. \citet{luSS94} find little evidence of
depletion in the MS, which would lower \C/\O\ but probably not \N/\O.

{\bf H1821+643:\ } The sightline to this quasar passes through the
warp of the outer Galaxy, seen in lower ionization species and \C\ at
$v<-$100~km~s\ts{-1} \citep{savage93b}.  The sightline also passes
near the HVC Complex C, and through the edge of radio Loop III.
\citet{SSL95} note that the sightline passes near planetary nebula
K1-16, but probably not near enough to contribute to the \N\ and \C\
absorption.  \citet{savage02} claim however that the planetary nebula
does contribute to \O.
\citet{SSL95} suggest that the bulk of the absorption, near 
0~km~s\ts{-1} LSR velocity, may be associated with the Loop III
superbubble.  They argue that the absorbing gas is hot based on the
large line widths and low \C/\N\ ratio, which is consistent with
collisional ionization equilibrium.  This is confirmed by an
anomalously high ratio log[N(\N)/N(\O)]$\sim -$0.1 observed here. The
\C/\O\ and \N/\O\ ratios found here are also consistent with
collisional ionization equilibrium (in contrast to almost all of the
other gas observed in this project), and in particular the high \N/\O\
and \C/\O\ ratios imply a temperature of T$\simeq$10\ts{5.3}~K.
\citet{SSL95} could not distinguish between that temperature and a
higher one (6--10\up{5}~K) based on their single ion ratio.
 
Absorption is detected in all 3 lines near $-$100~km~s\ts{-1}, with
somewhat low \N/\O\ and \C/\O\ column density ratios; this is
consistent with somewhat higher ionization state in the more extended
gas of the warp, but the gradient in ion ratio is not significant
enough to make a definitive claim.  \citet{SSL95} argued that this
gas, which they observed primarily in \C, is not related to the
Galactic fountain because the supernova rate and energy to drive the
fountain are low in the outer Galaxy. They go on to propose that the
gas could be photoionized. However, detection of \O\ at similar
velocities in this study and by \citet{oegerle00} argues for a
collisional ionization origin of the gas, and is thus left in the
general sample of this study.  Absorption detected near
$-$200~km~s\ts{-1} may be associated with Complex C, but the ion
column density ratios are not significantly deviant from average halo
values, and is more likely associated with the maximum velocity
allowed by the distant Galactic warp gas.  \citep[This gas was also
noted in \O\ absorption by][]{oegerle00}

{\bf Markarian 110:\ } The sightline to this Seyfert galaxy lies just
off the edge of HVC Complex A, and just outside of Galactic radio Loop
III.  This sightline was excluded from the analysis because of the low
quality of the FUSE data and the questionable \N\ detection, although
the sightline is probably acceptable for halo study with better data.

{\bf Markarian 116:\ } (I Zwicky 18) This sightline pierces high
velocity cloud Complex A and a marginal detection of \N\ at
$v\sim-$150~km~s\ts{-1} might be associated, but this is rejected as
is the entire sightline, because no low-velocity \N\ is detected.

{\bf Markarian 279:\ } The sightline to this Seyfert I galaxy passes
through HVC Complex C and the edge of radio Loop III.  Extended
negative velocity absorption $\lesssim -$100~km~s\ts{-1} seen in \N\
and \O\ could either be Galactic halo gas at velocities permitted by
the rotation curve or could be associated with the HVC.  As there is
no clear separation of the absorption into components, the entire
velocity range is kept in the sample. \citet{penton1} suggested that
the \N$\lambda$1239 line might be intergalactic Ly$\alpha$, but
because the $\lambda$1242 is definitely confused with intergalactic
absorption this identification can neither be confirmed nor denied,
and it is assumed to be \N\ here.  No indication that the radio Loop
is accounting for atypical conditions is evident in these
observations.

{\bf Markarian 290:\ } 
This sightline also passes through HVC Complex C, but probably
misses radio Loop III.  The FUSE data are only of modest quality, but
gas is clearly detected in \N\ and \O.  The column density ratio
\N/\O\ shows a large variation in the high velocity ($v\sim
-$100~km~s\ts{-1}) gas, but the error bars are large.  Complex C was
reported detected in \ion{Ca}{2} along this sightline, at
$v=-$137~km~s\ts{-1} \citep{wakker97}, close to the same velocity as
the \N\ and \O\ absorption.

{\bf Markarian 335:\ } This sightline just misses the tail of the
Magellanic Stream and passes through the center of Galactic radio
Loop II.  There is \O\ at large negative velocities which could be
associated with the MS ($-$350~$<v<-$150~km~s\ts{-1}). \N\ absorption
is extremely weak, and although there is the hint of a very low \N/\O\
ratio in both the low and high velocity gas, the ratio could not be
determined well enough to include in the sample.

{\bf Markarian 421:\ } The closest HVC to this BL Lac sightline is
Complex M, several degrees away.  \N\ column densities are low but not
undetected.


{\bf Markarian 509:\ } \label{mkn509} \citet{SSL97} report total
(low-velocity) \N\ and \C\ column densities towards Mkn 509 consistent
with those found here, and the ion ratios agree with those reported by
\citet{3c273}.  This sightline is particularly interesting because of
the detection of highly ionized high velocity gas, undetected in
neutral hydrogen; \citet{SSLM95} see \C\ at
$-$340~$<v<$~$-$170~km~s\ts{-1} and no \ion H1 emission at those
velocities (though there is \ion H1 emission at similar velocities 1-2
degrees away).  They only place an upper limit on \N, and argue that
the clouds may be photoionized because conductive interface and
cooling gas models have difficult producing the large ratios \C/\N\
$\gtrsim$5 they observe.  Photoionization does require the clouds to
be large (at least a kiloparsec).  In this dataset, high velocity gas
is observed in \O\ as well as \C, and again the \N\ detection is
marginal at best.  The presence of \O\ tends to argue against
photoionization, because photoionization leads to unreasonably large
cloud sizes if the absorbers are associated with our Galactic halo.
An ionization parameter $U = n_\gamma/n_H \gtrsim 10^{-1}$ would be
required to produce the observed column density ratio
log[N(\N)/N(\O)]$\lesssim$-1 via photoionization by the extragalactic
background of quasars and AGN \citep{tripp00}.  This translates into a
density upper limit $\log(n_H)\lesssim-$5.5,
$\log(n_{OVI})\lesssim-$9, and the observed \O\ column density yields
a physical cloud size of $R\gtrsim$100~kpc (although the \O\
absorption is confused with molecular hydrogen, assuming the maximum
2$\sigma$ allowed H$_2$ column from other H$_2$ lines still leaves
N(\O)$\gtrsim$1.2\up{14}~cm\ts{-2}).  Finally, it is not clear that
similar \C\ and \O\ column densities can be reconciled with very low
\N\ in a purely photoionization model.  The three ions are still
consistent with turbulent mixing layer models \citep[also noted
by][]{SSL97} but large numbers of such interfaces would be required to
produce the total column densities.


{\bf Markarian 817:\ } The sightline passes through HVC complex C.  As
with the other sightlines in this part of the sky, it is not possible
to determine whether the gas seen at negative velocities is part of
the extended disk or the high velocity cloud.

{\bf Markarian 876:\ } This sightline represents the first detection
of \O\ in a high velocity cloud, namely Complex C \citep{murphy00}.
Unfortunately the \N\ absorption is only marginally detected in the
high velocity gas.  The ion ratio in the high velocity gas is
log[N(\N)/N(\O)]$\sim-$0.8 in the HVC, not significantly different
from the range in the low velocity gas along this sightline or in the
halo in general.

{\bf Markarian 926:\ } This sightline just misses the tail end of the
Magellanic Stream, but does pass through Galactic radio Loop III.
The FUSE data was of insufficient quality to include this sightline in
further analysis.

{\bf Markarian 1383:\ } This sightline passes through Loop I and near Loop IV.
No unusual gas is detected in the line profiles or ion ratios.

{\bf Markarian 1502:\ } (I Zwicky 1) The sightline to this narrow-line
quasar passes through the center of radio Loop II.  Absorption seen
near 1031~\AA\ is probably highly contaminated by molecular hydrogen,
which is present along this sightline (although the modest
signal-to-noise does not permit a very accurate determination of the
column density), so only the low-velocity gas detected in \O\ and \N\
is analyzed.  The \N/\O\ ratio is fairly high
(log[N(\N)/N(\O)]$\sim-$0.6) but has large error bars, so the
difference does not merit special treatment of the sightline.

{\bf Markarian 1513:\ } (II Zwicky 136) This sightline passes through
Galactic radio Loop II, but no anomalies are detected in the column
densities profiles or their ratio.


{\bf NGC 3783:\ } The sightline to the nucleus of this Seyfert I
galaxy passes through the interior of Loop I and through the HVC
287.5+22.5+240.  \citet{sembach01} report on the FUSE observations of
this high velocity gas and argue that it is the leading edge of the
Magellanic Stream.  As they point out, any \O\ absorption in the high
velocity gas (high positive or negative velocities) is obscured by
intrinsic Ly$\beta$ absorption: NGC 3783 has a redshift of
2929$\pm$3~km~s\ts{-1} \citep{theureau98} and prominent absorbers at
$-$560 and $-$1420~km~s\ts{-1} \citep{crenshaw99}.  The current data
show no definite \C\ or \N\ absorption at
$|v|\gtrsim$100~km~s\ts{-1}, strengthening the claim that little of
the observed absorption near 1031~\AA\ is \O.

{\bf NGC 4151:\ } The sightline towards this nearly Seyfert I galaxy
does not traverse any significant Galactic features, but isolation of
the Galactic absorption does require some care because of absorption
intrinsic to that galaxy.  NGC 4151 has a variable outflow of hot gas
(including absorption by the Li-like ions) at 0 to 1600~km~s\ts{-1}
\citep[the Seyfert has a redshift of
$\simeq$1000~km~s\ts{-1},][]{weymann97}.  \citet{weymann97} fit the
Galactic absorption from \C\ together with the intrinsic, while our
study treats the broad intrinsic absorption as a complex
continuum. Comparison of the results suggests that the Galactic \C\
absorption reported here is an underestimate because of the
line-blanketing effect of intrinsic absorption.  The line ratios
\N/\O\ and \C/\O\ found here are not particularly different from the
halo mean.

{\bf NGC 5548:\ } The integrated values for \N\ and \C\ column
densities are in agreement with \citet{SSL97} along the unremarkable
sightline towards this Seyfert I galaxy.

{\bf PG0804+761:\ } The sightline towards this low-redshift quasar
passes through the center of radio Loop III, very near high velocity
cloud Complex A, and over the outer Galaxy spiral arms.  Although
there is \O\ emission at moderately high velocities, there is none at
$-$180~km~s\ts{-1}, where Complex A gas would be expected
\citep{richter01a}.  The \N\ absorption at that velocity is obscured by
what appears to be intergalactic Ly$\alpha$ absorption.  At lower
velocities (data in the wings of the Ly$\alpha$ absorber is
discarded) the \N\ is weak but present.  There is a break in the
\N/\O\ ratio near $-$40~km~s\ts{-1} which might indicate a different
ratio in low halo gas compared to the low-latitude
Intermediate-Velocity Arch \citep[see][for further description of the
intermediate-velocity gas along this sightline.]{richter01a}

{\bf PG0953+414:\ } The nearest high-velocity gas to the sightline
towards this low-redshift QSO is 1.7\dg\ away at $-$91~km~s\ts{-1}
\citep{savagehst00}.  No anomalies are detected.

{\bf PG1116+215:\ } Although there is no apparent \ion H1
high-velocity gas near the sightline towards this low-redshift quasar,
\citet{tripp98} and \citet{savagehst00} detect absorption from
modestly ionized species (\ion{Si}{2}, \ion{Mg}{2}, \ion C2,
\ion{Si}{4}) at $v\sim$200~km~s\ts{-1}.  The detection is confirmed
here in the Li-like ions, with ion ratios log[N(\N)/N(\O)]$\sim-$0.7
and log[N(\N)/N(\O)]$\sim-$0.1.


{\bf PG1259+593:\ } The sightline to the quasar PG1259+593 passes
through the Intermediate-Velocity (IV) Arch ($-$55~km~s\ts{-1}), and
high-velocity cloud Complex C ($-$130~km~s\ts{-1}).  The detection of
\O\ in these features reported by \citet{richter01b} is confirmed here,
as well as marginal detection of \N\ with log[N(\N)/N(\O)]$\sim-$0.9.
\citet{savage93b} note negative extensions of the absorption from \C\
and lower ions which may be associated with IV gas, but the Faint
Object Spectrograph resolution is insufficient to properly resolve
components.

{\bf PG1351+640:\ } The sightline towards this broad-absorption-line
QSO passes through the high velocity cloud Complex C, but the FUSE
data quality is insufficient to accurately study the Galactic
absorption \citep[see][for analysis of the intrinsic absorbers in the
FUSE spectrum]{zheng02}.


{\bf PKS 2155-304:\ } The sightline to this BL Lac does not pass
through any apparent high velocity gas detected in \ion H1 emission,
but it does show high velocity absorption in \ion{O}{6}, \ion{C}{4}, 
\ion{Si}{4}, and certain low ions \citep{collins04}.  Although these 
HVCs have been attributed to gas in the Local Group, the presence 
of the low ions (\ion{S}{2}, \ion{C}{2}) is more consistent with
higher-density gas in the Galactic halo.  Integrated column densities 
for the low-velocity halo gas
agree with \citet{SSL97} (\N\ and \C) and \citet{3c273} (\N/\O\ and
\C/\O; the value for the latter is $\sim 1\sigma$ higher than their
value).  The \N$\lambda$1239 line is a questionable identification; 
the above two papers associate it with Galactic \N, but
\citet{shull98} associate it with intergalactic Ly$\alpha$ because
of the weakness of the 1242~\AA\ line.
Because \N$\lambda$1242 is quite weak, we exclude this sightline
from general analysis of the N(\N)/N(\O) line ratio as a function of velocity.

{\bf RXJ 1230.8+0115:\ } Like the sightline towards 3C273, this
sightline passes through the edges of radio Loops I and IV.  The
sightline is not usable for general analysis because the \O\ line is
obscured by intergalactic Ly$\beta$.

{\bf TON S180:\ } This sightline lies about 10\dg\ off of the
Magellanic Stream, and \O\ has been previously detected at both high
positive and negative velocities \citep{sembach00}.  The positive
velocity gas is not detected in \N, and although there is a hint of
the negative-velocity gas, it is obscured by what is apparently
intergalactic Ly$\alpha$ absorption.

{\bf VII Zwicky 118:\ } This sightline passes through the center of
ratio Loop III, but no anomalies are detected in the column densities
profiles of \O\ and \ion C4.  Signal-to-noise is too low for \N\ to include 
in ratio analysis.

\clearpage
\begin{figure}[h]
\epsscale{0.8}
\plotone{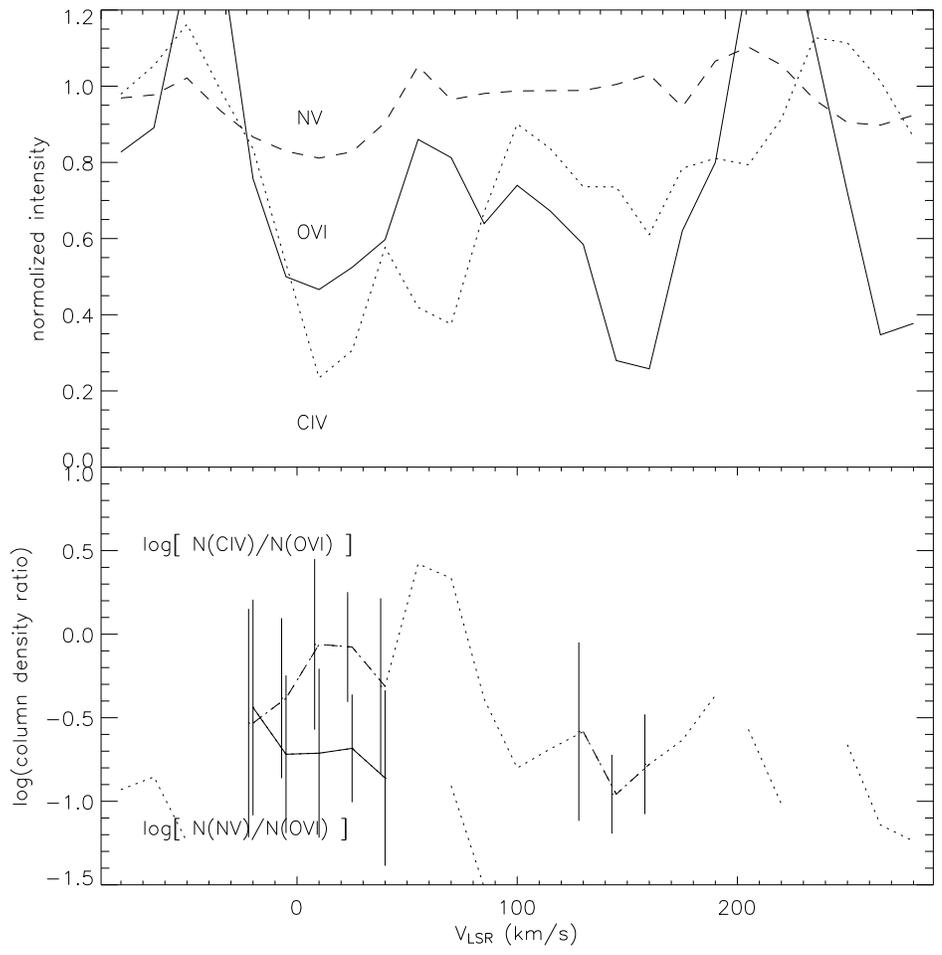}
\caption{
Fairall 9.  Lines as in Figure \ref{sightlines}. }
\end{figure}
\clearpage

\clearpage
\begin{figure}[h]
\plotone{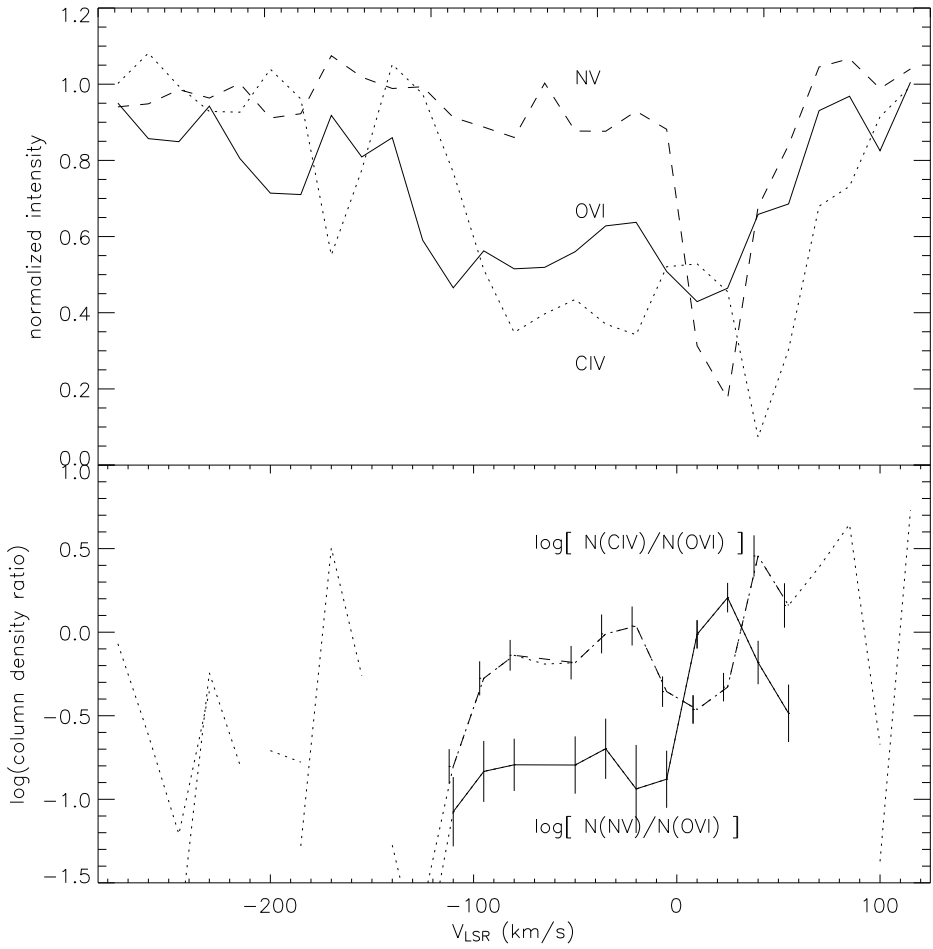}
\caption{
H1821+643.  Lines as in Figure \ref{sightlines}. }
\end{figure}
\clearpage

\clearpage
\begin{figure}[h]
\plotone{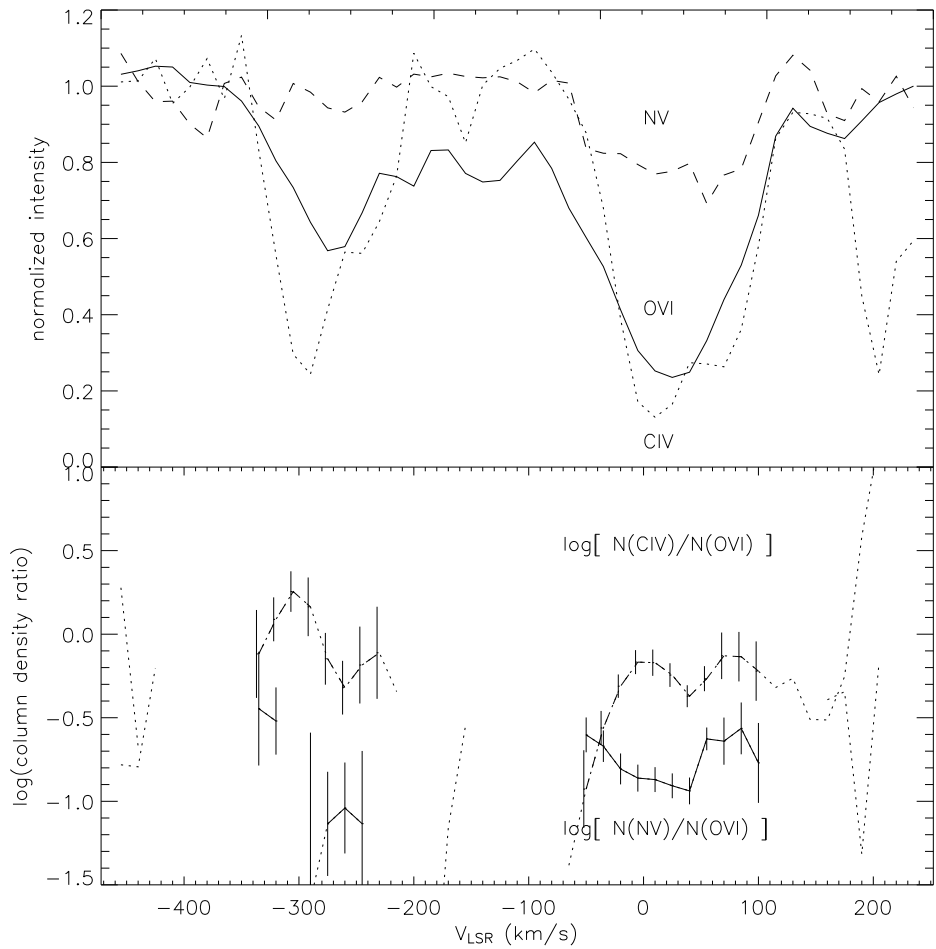}
\caption{
Markarian 509.  Lines as in Figure \ref{sightlines}. }
\end{figure}
\clearpage

\clearpage
\begin{figure}[h]
\plotone{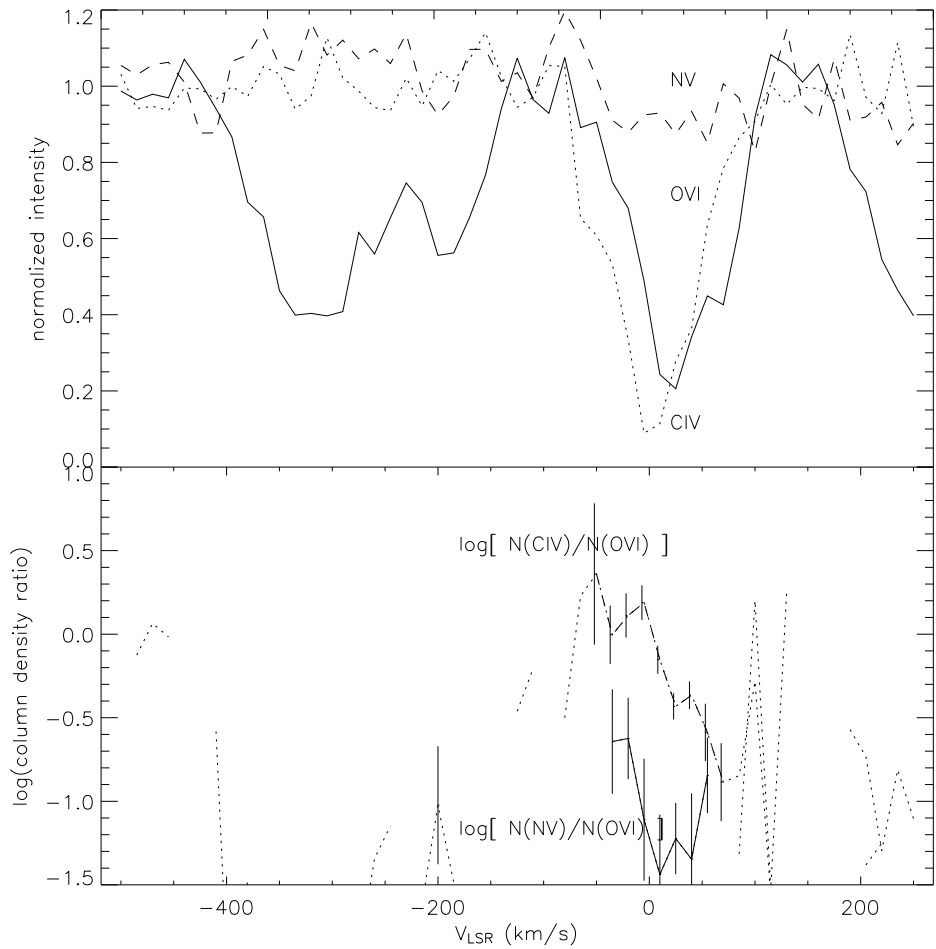}
\caption{
NGC3783.  Lines as in Figure \ref{sightlines}. }
\end{figure}
\clearpage

\clearpage
\begin{figure}[h]
\plotone{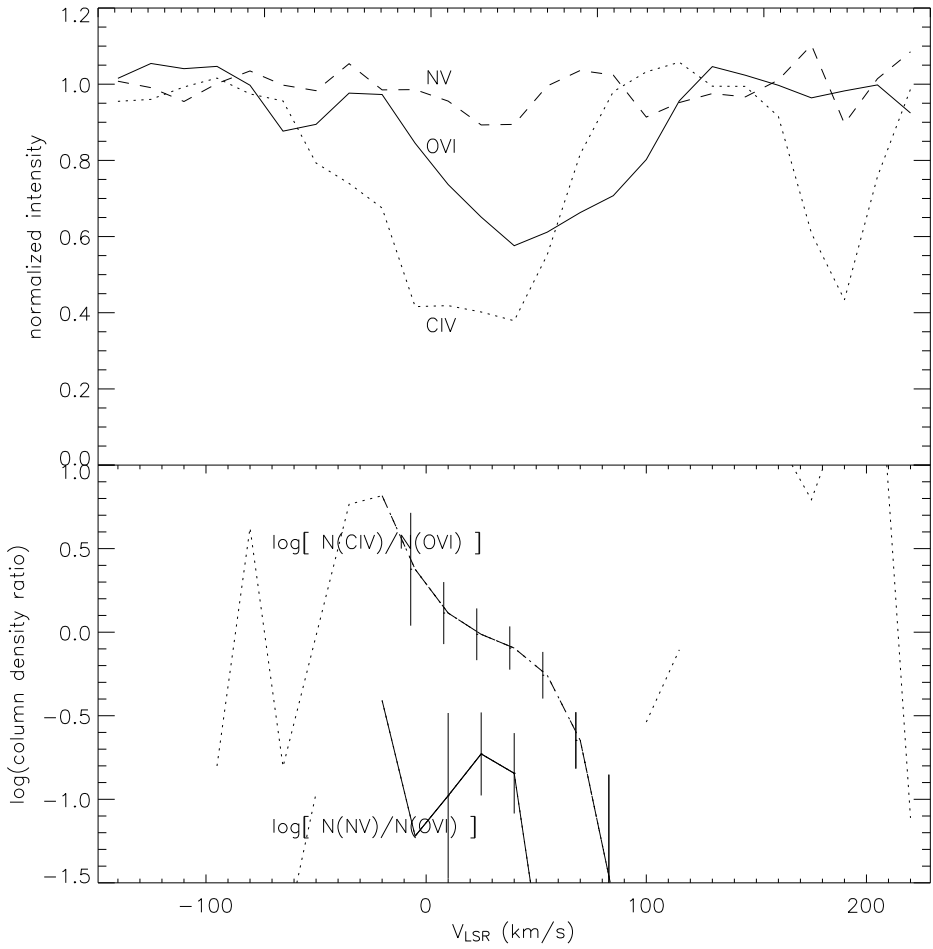}
\caption{
NGC4151.  Lines as in Figure \ref{sightlines}. }
\end{figure}
\clearpage

\clearpage
\begin{figure}[h]
\plotone{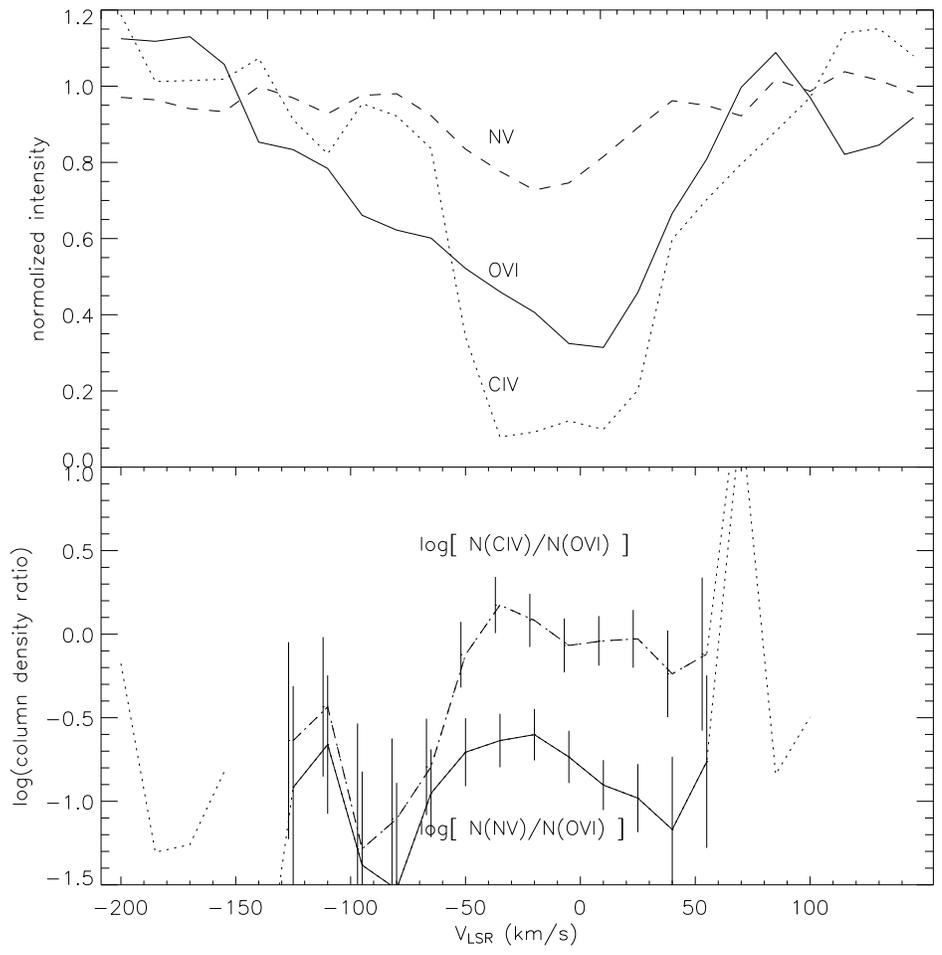}
\caption{
NGC5548.  Lines as in Figure \ref{sightlines}. }
\end{figure}
\clearpage

\clearpage
\begin{figure}[h]
\plotone{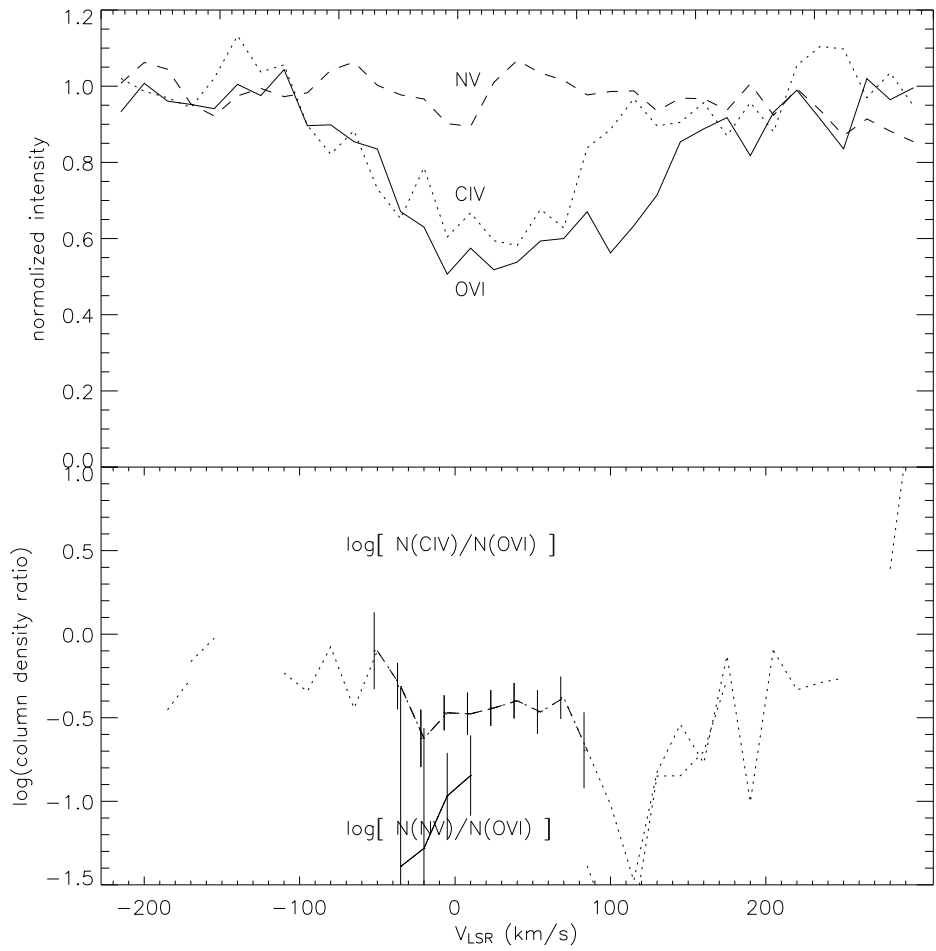}
\caption{
PG0953+414.  Lines as in Figure \ref{sightlines}. }
\end{figure}
\clearpage

\clearpage
\begin{figure}[h]
\plotone{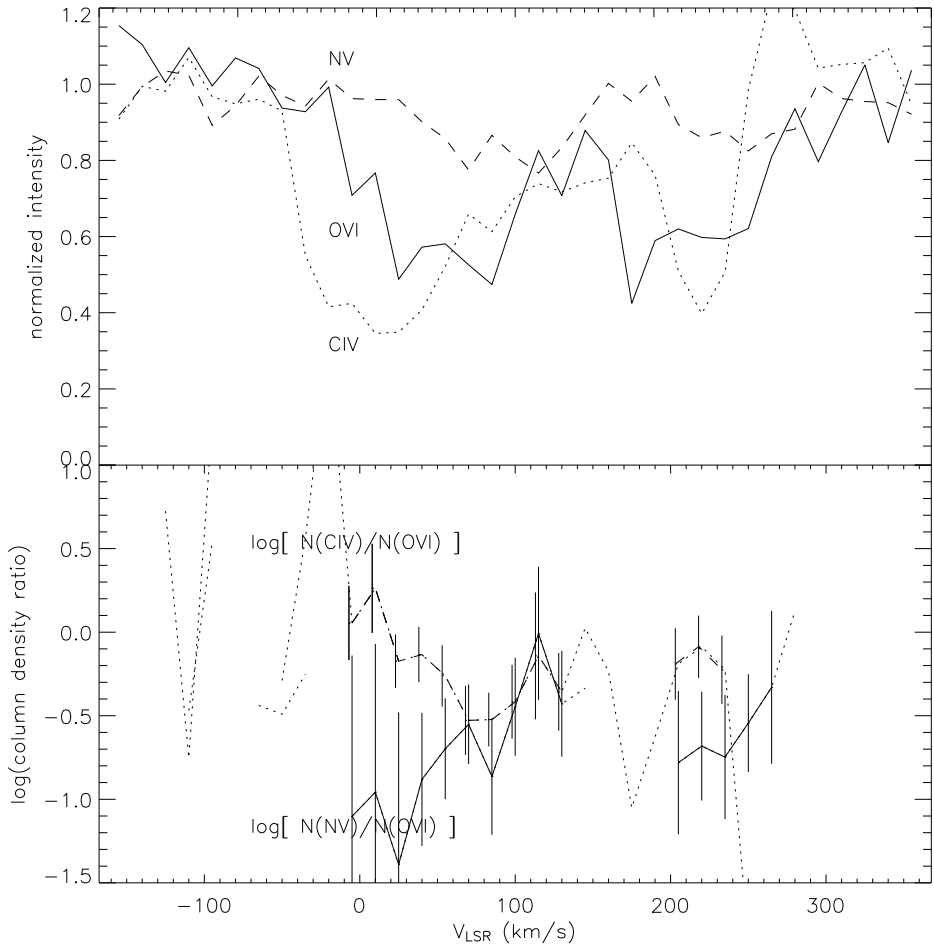}
\caption{
PG1116+215.  Lines as in Figure \ref{sightlines}. }
\end{figure}


\clearpage


\footnotesize
\begin{thebibliography}{129}
\expandafter\ifx\csname natexlab\endcsname\relax\def\natexlab#1{#1}\fi


\bibitem[{Benjamin \& Shapiro(1993)}]{benjamin93}
Benjamin, R., \& Shapiro, P. 1993, in Ultraviolet and {X}-ray Spectroscopy of
  Laboratory and Astrophysical Plasmas, ed. E.~Silver \& S.~Kahn (New York:
  Cambridge Univ. Press), 280


\bibitem[{{Berkhuijsen} {et~al.}(1971){Berkhuijsen}, {Haslam}, \&
  {Salter}}]{berk71}
{Berkhuijsen}, E.~M., {Haslam}, C. G.~T., \& {Salter}, C.~J. 1971, \aap, 14,
  252

\bibitem[{{B{\"o}hringer} \& {Hartquist}(1987)}]{bohringer87}
{B{\"o}hringer}, H., \& {Hartquist}, T.~W. 1987, \mnras, 228, 915

\bibitem[{{Borkowski} {et~al.}(1990){Borkowski}, {Balbus}, \&
  {Fristrom}}]{borkowski90}
{Borkowski}, K.~J., {Balbus}, S.~A., \& {Fristrom}, C.~C. 1990, \apj, 355, 501

\bibitem[{{Brandt} {et~al.}(1994)}]{ghrs}
{Brandt}, J.~C., {et~al.} 1994, \pasp, 106, 890

\bibitem[{{Bregman}(1980)}]{bregman80}
{Bregman}, J.~N. 1980, \apj, 236, 577

\bibitem[{Burton \& Hartmann(1994)}]{LDS:proc}
Burton, W.~B., \& Hartmann, D. 1994, in ASP Conference Series, Vol.~67,
  Unveiling large-scale structures behind the {M}ilky {W}ay, ed. C.~Balkowski
   R.~C. Kraan-Korteweg, 31

\bibitem[{Collins {et~al.}(2003)Collins, Shull, \& Giroux}]{collins03}
Collins, J., Shull, J.~M., \& Giroux, M.~L. 2003, \apj, 585, 336 

\bibitem[{Collins {et~al.}(2004)Collins, Shull, \& Giroux}]{collins04}
Collins, J., Shull, J.~M., \& Giroux, M.~L. 2004, \apj, 605, in press 

\bibitem[{{Crenshaw} {et~al.}(1999){Crenshaw}, {Kraemer}, {Boggess}, {Maran},
  {Mushotzky}, \& {Wu}}]{crenshaw99}
{Crenshaw}, D.~M., {Kraemer}, S.~B., {Boggess}, A., {Maran}, S.~P.,
  {Mushotzky}, R.~F., \& {Wu}, C. 1999, \apj, 516, 750

\bibitem[{Davies {et~al.}(2001)}]{stisweb}
Davies, J., {et~al.} 2001, {STIS} Instrument Handbook, available electronically
  from http://www.stsci.edu/cgi-bin/stis?cat=documents{\&}subcat=ihb

\bibitem[{{Dixon} {et~al.}(2001){Dixon}, {Sallmen}, {Hurwitz}, \&
  {Lieu}}]{dixon01}
{Dixon}, W.~V.~D., {Sallmen}, S., {Hurwitz}, M., \& {Lieu}, R. 2001, \apjl,
  552, L69

\bibitem[{{Gibson} {et~al.}(2001){Gibson}, {Giroux}, {Penton}, {Stocke},
  {Shull}, \& {Tumlinson}}]{gibson01}
{Gibson}, B.~K., {Giroux}, M.~L., {Penton}, S.~V., {Stocke}, J.~T., {Shull},
  J.~M., \& {Tumlinson}, J. 2001, \aj, 122, 3280

\bibitem[{Haynes {et~al.}(1999)}]{parkes}
Haynes, R., {et~al.} 1999, in New Views of the Magellanic Clouds, ed. Y.-H. Chu
  {et~al.}, ASP, San Francisco, 63

\bibitem[{{Heiles} {et~al.}(1980){Heiles}, {Chu}, {Troland}, {Reynolds}, \&
  {Yegingil}}]{heiles80}
{Heiles}, C., {Chu}, Y., {Troland}, T.~H., {Reynolds}, R.~J., \& {Yegingil}, I.
  1980, \apj, 242, 533

\bibitem[{{Hou} {et~al.}(2000){Hou}, {Prantzos}, \& {Boissier}}]{hou01}
{Hou}, J.~L., {Prantzos}, N., \& {Boissier}, S. 2000, \aap, 362, 921

\bibitem[{{Howk} {et~al.}(2002){Howk}, {Savage}, {Sembach}, \&
  {Hoopes}}]{howk02}
{Howk}, J.~C., {Savage}, B.~D., {Sembach}, K.~R., \& {Hoopes}, C.~G. 2002,
  \apj, 572, 264

\bibitem[{Indebetouw \& Shull(2004)}]{remy04}
{Indebetouw}, R., \& {Shull}, J. M. 2004, \apj, 605, in press (Paper I) 

\bibitem[{{Iwan}(1980)}]{iwan80}
{Iwan}, D. 1980, \apj, 239, 316

\bibitem[{Jenkins(2001)}]{jenkinsparis}
Jenkins, E. 2001, in Gaseous Matter in Interstellar and Intergalactic Space,
  IAP Colloquium 17, ed. R.~Ferlet, M.~Lemoine, J.-M. Desert, B.~Raban,
  (Paris: Frontier Group), 99

\bibitem[{{Jenkins}(1996)}]{jenkins96}
{Jenkins}, E.~B. 1996, \apj, 471, 292

\bibitem[{{Kimble} {et~al.}(1998)}]{stis}
{Kimble}, R.~A., {et~al.} 1998, \apjl, 492, L83

\bibitem[{{Lu} {et~al.}(1994){Lu}, {Savage}, \& {Sembach}}]{luSS94}
{Lu}, L., {Savage}, B.~D., \& {Sembach}, K.~R. 1994, \apj, 437, L119

\bibitem[{{Moos} {et~al.}(2000)}]{fuse1}
{Moos}, H.~W., {et~al.} 2000, \apjl, 538, L1

\bibitem[{{Morras}(1983)}]{morras83}
{Morras}, R. 1983, \aj, 88, 62

\bibitem[{{Murphy} {et~al.}(2000)}]{murphy00}
{Murphy}, E.~M., {et~al.} 2000, \apjl, 538, L35

\bibitem[{Oegerle {et~al.}(2000a)Oegerle, Murphy, \& Kriss}]{fusedata}
Oegerle, W. R., Murphy, E., \& Kriss, J. 2000a, The FUSE Data Handbook, available
  electronically from http://fuse.pha.jhu.edu/analysis/dhbook.html

\bibitem[{{Oegerle} {et~al.}(2000b)}]{oegerle00}
{Oegerle}, W.~R. {et~al.} 2000b, \apjl, 538, L23

\bibitem[{{Penton} {et~al.}(2000{\natexlab{a}}){Penton}, {Shull}, \&
  {Stocke}}]{penton2}
{Penton}, S.~V., {Shull}, J.~M., \& {Stocke}, J.~T. 2000{\natexlab{a}}, \apj,
  544, 150

\bibitem[{{Penton} {et~al.}(2000{\natexlab{b}}){Penton}, {Stocke}, \&
  {Shull}}]{penton1}
{Penton}, S.~V., {Stocke}, J.~T., \& {Shull}, J.~M. 2000{\natexlab{b}}, \apjs,
  130, 121

\bibitem[{{Penton} {et~al.}(2004){Penton}, {Stocke}, \& {Shull}}]{penton3}
---. 2004, \apjs, 132, in press

\bibitem[{{Richter} {et~al.}(2001a){Richter}, {Savage}, {Wakker}, {Sembach}, \&
  {Kalberla}}]{richter01a}
{Richter}, P., {Savage}, B.~D., {Wakker}, B.~P., {Sembach}, K.~R., \&
  {Kalberla}, P. M.~W. 2001a, \apj, 549, 281

\bibitem[{Richter {et~al.}(2001b)Richter, Sembach, Wakker, Savage, Tripp,
  Murphy, Kalberla, Jenkins}]{richter01b}
Richter, P., et al. 2001b, \apj, 559, 318

\bibitem[{{Rolleston} {et~al.}(2000){Rolleston}, {Smartt}, {Dufton}, \&
  {Ryans}}]{rolleston00}
{Rolleston}, W. R.~J., {Smartt}, S.~J., {Dufton}, P.~L., \& {Ryans}, R. S.~I.
  2000, \aap, 363, 537

\bibitem[{{Sahnow} {et~al.}(2000)}]{sahnow00}
{Sahnow}, D.~J., {et~al.} 2000, \apjl, 538, L7

\bibitem[{Savage(2001)}]{savageparis}
Savage, B. 2001, in Gaseous Matter in Interstellar and Intergalactic Space, IAP
  Colloquium 17, ed. R.~Ferlet, M.~Lemoine, J.-M. Desert, B.~Raban, 
  (Paris: Frontier Group), 109

\bibitem[{{Savage} {et~al.}(1993{\natexlab{a}}){Savage}, {Lu}, {Weymann},
  {Morris}, \& {Gilliland}}]{savage93}
{Savage}, B.~D., {Lu}, L., {Weymann}, R.~J., {Morris}, S.~L., \& {Gilliland},
  R.~L. 1993{\natexlab{a}}, \apj, 404, 124

\bibitem[{{Savage} \& {Sembach}(1991)}]{ss91}
{Savage}, B.~D., \& {Sembach}, K.~R. 1991, \apj, 379, 245

\bibitem[{{Savage} {et~al.}(1995){Savage}, {Sembach}, \& {Lu}}]{SSL95}
{Savage}, B.~D., {Sembach}, K.~R., \& {Lu}, L. 1995, \apj, 449, 145

\bibitem[{{Savage} {et~al.}(1997){Savage}, {Sembach}, \& {Lu}}]{SSL97}
---. 1997, \aj, 113, 2158

\bibitem[{{Savage} {et~al.}(1993{\natexlab{b}})}]{savage93b}
{Savage}, B.~D., {et~al.} 1993{\natexlab{b}}, \apj, 413, 116

\bibitem[{{Savage} {et~al.}(2000{\natexlab{a}})}]{fuse2}
---. 2000{\natexlab{a}}, \apjl, 538, L27

\bibitem[{{Savage} {et~al.}(2000{\natexlab{b}})}]{savagehst00}
---. 2000{\natexlab{b}}, \apjs, 129, 563

\bibitem[{Savage {et~al.}(2003)}]{savage02}
Savage, B.~D., {et~al.} 2003, \apjs, 146, 125

\bibitem[{{Sembach} {et~al.}(2001{\natexlab{a}}){Sembach}, {Howk}, {Savage}, \&
  {Shull}}]{sembach01}
{Sembach}, K.~R., {Howk}, J.~C., {Savage}, B.~D., \& {Shull}, J.~M.
  2001{\natexlab{a}}, \aj, 121, 992

\bibitem[{{Sembach} {et~al.}(2001{\natexlab{b}}){Sembach}, {Howk}, {Savage},
  {Shull}, \& {Oegerle}}]{3c273}
{Sembach}, K.~R., {Howk}, J.~C., {Savage}, B.~D., {Shull}, J.~M., \& {Oegerle},
  W.~R. 2001{\natexlab{b}}, \apj, 561, 573

\bibitem[{{Sembach} \& {Savage}(1992)}]{SS92}
{Sembach}, K.~R., \& {Savage}, B.~D. 1992, \apjs, 83, 147

\bibitem[{{Sembach} {et~al.}(1999){Sembach}, {Savage}, \& {Hurwitz}}]{SSH99}
{Sembach}, K.~R., {Savage}, B.~D., \& {Hurwitz}, M. 1999, \apj, 524, 98

\bibitem[{{Sembach} {et~al.}(1995){Sembach}, {Savage}, {Lu}, \&
  {Murphy}}]{SSLM95}
{Sembach}, K.~R., {Savage}, B.~D., {Lu}, L., \& {Murphy}, E.~M. 1995, \apj,
  451, 616

\bibitem[{{Sembach} {et~al.}(2000)}]{sembach00}
{Sembach}, K.~R., {et~al.} 2000, \apjl, 538, L31

\bibitem[{Sembach {et~al.}(2003)}]{sembach02}
Sembach, K.~R., {et~al.} 2003, \apjs, 146, 165

\bibitem[{Shapiro \& Benjamin(1993)}]{shapiro93}
Shapiro, P., \& Benjamin, R. 1993, in Star Formation, Galaxies and the
  Interstellar Medium, ed. J.~Franco, F.~Ferrini, G.~Tenorio-Tagle (New
  York: Cambridge Univ. Press), 275

\bibitem[{{Shelton} {et~al.}(2001)}]{shelton01}
{Shelton}, R.~L., {et~al.} 2001, \apj, 560, 730

\bibitem[{{Shelton}(1998)}]{shelton98}
{Shelton}, R.~L. 1998, \apj, 504, 785

\bibitem[{{Shull} {et~al.}(1998){Shull}, {Penton}, {Stocke}, {Giroux}, {van
  Gorkom}, {Lee}, \& {Carilli}}]{shull98}
{Shull}, J.~M., {Penton}, S.~V., {Stocke}, J.~T., {Giroux}, M.~L., {van
  Gorkom}, J.~H., {Lee}, Y.~H., \& {Carilli}, C. 1998, \aj, 116, 2094

\bibitem[{{Shull} \& {Slavin}(1994)}]{shullslavin94}
{Shull}, J.~M., \& {Slavin}, J.~D. 1994, \apj, 427, 784

\bibitem[{{Shull} {et~al.}(2000)}]{shull2000}
{Shull}, J.~M., {et~al.} 2000, \apjl, 538, L73

\bibitem[{{Slavin} \& {Cox}(1993)}]{slavincox93}
{Slavin}, J.~D., \& {Cox}, D.~P. 1993, \apj, 417, 187.

\bibitem[{{Slavin} {et~al.}(1993){Slavin}, {Shull}, \& {Begelman}}]{mixing}
{Slavin}, J.~D., {Shull}, J.~M., \& {Begelman}, M.~C. 1993, \apj, 407, 83

\bibitem[{{Snowden} {et~al.}(1995)}]{snowden95}
{Snowden}, S.~L., {et~al.} 1995, \apj, 454, 643

\bibitem[{Soderblom {et~al.}(1995)Soderblom, Gonnella, Hulbert, Leitherer,
  Schultz, \& Sherbert}]{ghrsweb}
Soderblom, D., Gonnella, A., Hulbert, S., Leitherer, C., Schultz, A., \&
  Sherbert, L. 1995, {GHRS} Instrument Handbook, available electronically from
  http://www.stsci.edu/instrument-news/handbooks/ghrs/GHRS{\_}1.html

\bibitem[{{Sutherland} \& {Dopita}(1993)}]{sutherland93}
{Sutherland}, R.~S., \& {Dopita}, M.~A. 1993, \apjs, 88, 253

\bibitem[{{Theureau} {et~al.}(1998){Theureau}, {Bottinelli}, {Coudreau-Durand},
  {Gouguenheim}, {Hallet}, {Loulergue}, {Paturel}, \&
  {Teerikorpi}}]{theureau98}
  {Theureau}, G., {Bottinelli}, L., {Coudreau-Durand}, N., {Gouguenheim}, L.,
  {Hallet}, N., {Loulergue}, M., {Paturel}, G., \& {Teerikorpi}, P. 1998,
  \aaps, 130, 333

\bibitem[{{Tripp} {et~al.}(1998){Tripp}, {Lu}, \& {Savage}}]{tripp98}
{Tripp}, T.~M., {Lu}, L., \& {Savage}, B.~D. 1998, \apj, 508, 200

\bibitem[{{Tripp} \& {Savage}(2000)}]{tripp00}
{Tripp}, T.~M., \& {Savage}, B.~D. 2000, \apj, 542, 42

\bibitem[{{Tripp} {et~al.}(1993){Tripp}, {Sembach}, \& {Savage}}]{trippSS93}
{Tripp}, T.~M., {Sembach}, K.~R., \& {Savage}, B.~D. 1993, \apj, 415, 652

\bibitem[{Wakker {et~al.}(2003)Wakker, Savage, Sembach, Richter, Meade, \&
  Jenkins}]{wakker02}
  Wakker, B.~P., Savage, B.~D., Sembach, K.~R., Richter, P., Meade, M., \&
  Jenkins, E.~B. 2003, \apjs, 146, 1

\bibitem[{{Wakker} \& {van Woerden}(1991)}]{wakker91}
{Wakker}, B.~P., \& {van Woerden}, H. 1991, \aap, 250, 509

\bibitem[{{Wakker} \& {van Woerden}(1997)}]{wakker97}
---. 1997, \araa, 35, 217

\bibitem[{{Wakker} {et~al.}(1999)}]{wakker99}
{Wakker}, B.~P., {et~al.} 1999, Nature, 402, 388

\bibitem[{{Weymann} {et~al.}(1997){Weymann}, {Morris}, {Gray}, \&
  {Hutchings}}]{weymann97}
{Weymann}, R.~J., {Morris}, S.~L., {Gray}, M.~E., \& {Hutchings}, J.~B. 1997,
  \apj, 483, 717

\bibitem[{Zheng {et~al.}(2001)}]{zheng02}
Zheng, W., {et~al.} 2001, \apj, 562, 152

\end{thebibliography}
\end{document}